\pdfoutput=1
\documentclass[11pt]{article}
\usepackage{geometry}
\usepackage{amsfonts}
\usepackage{amsmath,amssymb,amsthm}
\usepackage{mathrsfs}
\usepackage{float}
\usepackage{mathtools}
\usepackage{wasysym}

\usepackage[utf8]{inputenc}
\usepackage{caption,xspace}
\usepackage{subcaption}
\numberwithin{equation}{section}

\usepackage{graphicx}

\usepackage{soul}
\usepackage[sort&compress,numbers,colon]{natbib}
\usepackage{multirow}
\usepackage{color}
\usepackage[colorlinks]{hyperref}
\hypersetup{
pdfauthor={Tony Gherghetta, Benedict von Harling, Anibal D. Medina and Michael A. Schmidt},
pdftitle={The price of being SM-like in SUSY}
}

\usepackage{verbatim}
\newcommand{\be}{\begin{equation}}
\newcommand{\ee}{\end{equation}}
\newcommand{\bea}{\begin{eqnarray}}
\newcommand{\eea}{\end{eqnarray}}

\def\4vol{{\int d^4x \sqrt{-g}}}

\def\beq{\begin{equation}}
\def\eeq{\end{equation}}
\def\bea{\begin{eqnarray}}
\def\eea{\end{eqnarray}}
\def\bitem{\begin{itemize}}
\def\eitem{\end{itemize}}

\newcommand{\Eqref}[1]{eq.~(\ref{#1})}

\newcommand{\Secref}[1]{sec.~\ref{#1}}

\newcommand{\GeV}{\ensuremath{\, \mathrm{GeV}}}
\newcommand{\TeV}{\ensuremath{\, \mathrm{TeV}}}

\newcommand{\nc}{\newcommand}

\nc{\nt}{\tilde{N}}
\nc{\ra}{\rightarrow}
\nc{\lsim}{\begin{array}{c}\,\sim\vspace{-21pt}\\< \end{array}}
\nc{\gsim}{\begin{array}{c}\sim\vspace{-21pt}\\> \end{array}}
\nc{\tnt}{\tilde{N}}
\nc{\tst}{\tilde{t}}

\nc{\LL}{L}
\nc{\vv}{\tilde{v}}

\title{}

\begin{document}
\allowdisplaybreaks[1]

\begin{titlepage}

\begin{center}
{\Large\textbf{The price of being SM-like in SUSY}}
\\[10mm]
{\large
Tony Gherghetta$^{a,b,}$\footnote{\texttt{tgher@umn.edu}},
Benedict von Harling$^{c,}$\footnote{\texttt{bharling@sissa.it}},
Anibal D. Medina$^{b,}$\footnote{\texttt{anibal.medina@unimelb.edu.au}},
Michael A.~Schmidt$^{b,}$\footnote{\texttt{michael.schmidt@unimelb.edu.au}}}
\\[5mm]
{\small\textit{
$^a$School of Physics \& Astronomy, University of Minnesota, Minneapolis, MN 55455, USA\\
$^b$ARC Centre of Excellence for Particle Physics at the Terascale,\\
School of Physics, The University of Melbourne, Victoria 3010, Australia\\
$^c$SISSA, via Bonomea 265, 34136 Trieste, Italy 
}
}
\end{center}

\vspace*{1.0cm}
\date{\today}

\begin{abstract}
\noindent
We compute the tuning in supersymmetric models associated with the constraints from collider measurements of the Higgs couplings to fermions and  
gauge bosons. In supersymmetric models, a $CP$-even state with SM Higgs couplings mixes with additional, heavier $CP$-even states, causing deviations in the Higgs couplings from SM values. These deviations are reduced as the heavy states are decoupled with large soft masses, 
thereby exacerbating the tuning associated with the electroweak scale. This new source of tuning is different from that derived from collider limits on stops, gluinos and Higgsinos. It can be offset with large $\tan\beta$ in the MSSM, however this compensating effect is limited in the NMSSM with a large Higgs-singlet coupling due to restrictions on large $\tan\beta$ from electroweak precision tests. We derive a lower bound on this tuning and show that the level of precision of Higgs coupling measurements at the LHC will probe naturalness in the NMSSM at the few-percent level. This is comparable to the tuning derived from superpartner limits in models with a low messenger scale and split families. Instead the significant improvement in sensitivity of Higgs coupling measurements at the ILC will allow naturalness in these models to be constrained at the per-mille level, beyond any tuning derived from direct superpartner limits.
\end{abstract}

\end{titlepage}

\setcounter{footnote}{0}



\section{Introduction}
The discovery of the Higgs boson at the Large Hadron Collider (LHC) with a mass near ${126 \GeV}$ confirms that the Higgs mechanism is responsible for electroweak symmetry breaking in the Standard Model (SM). However, the question of whether the Higgs boson is naturally light compared to the Planck scale or fine-tuned remains to be established. Supersymmetry provides a well-known natural solution to this hierarchy problem. The conspicuously absent superpartners from Run I at the LHC, however, have led to increasingly stringent limits on their masses. This makes an increased residual tuning among the parameters of the supersymmetric models necessary in order to obtain a vacuum expectation value (vev) at the electroweak scale. In the best-case scenario of the Next-to-Minimal Supersymmetric Standard Model (NMSSM) with Higgs-singlet coupling $\lambda$ near one, low messenger scale ($20 \TeV$) and split sparticle spectrum, the fine-tuning is at the 5$\%$ level \cite{Gherghetta:2012gb}.

In addition to direct limits on superpartner masses, the measurement of the Higgs couplings to fermions and gauge bosons provides another test of naturalness. For example, stops affect the Higgs couplings to photons and gluons at the one-loop level. Limits on the deviations of these couplings from the SM yield limits on the stop masses, thereby constraining naturalness \cite{Arvanitaki:2011ck,Craig:2013xia,Farina:2013ssa,Fan:2014txa}. Here we will instead consider the effect on the Higgs couplings from the additional $CP$-even states in the Higgs sector. This arises already at tree-level and is therefore potentially larger than the aforementioned one-loop effect. How these fields affect the Higgs couplings is best understood in a field basis where only one linear combination of Higgs doublets obtains a vev, which couples to SM particles precisely like the SM Higgs. In general, however, the particle which we identify with the $126 \GeV$ Higgs observed at the LHC is an admixture of this state with the other $CP$-even states in the Higgs sector (one in the Minimal Supersymmetric Standard Model (MSSM) and two in the NMSSM). This drives the Higgs couplings away from the SM values. The admixture arises from a non-diagonal mass matrix, and thus the deviations of the Higgs couplings from the SM can be made smaller in two different ways: On the one hand, the off-diagonal elements of the mass matrix can be made smaller. However they can typically be arbitrarily small only if there is an accidental cancellation among the various parameters that determine their value.
This is a new type of tuning that should be taken into account when assessing the naturalness of the model. Alternatively, the diagonal elements of the mass matrix corresponding to the additional Higgs states can be made larger. However, this requires increasing the soft masses which determine these diagonal elements. This exacerbates the hierarchy between the electroweak and the supersymmetry-breaking scale and thereby increases the fine-tuning. Therefore in either case, the closer the Higgs couplings become to those in the SM, the less natural the theory, and a tuning price must be paid for SM-like couplings in supersymmetric models.

The tuning associated with the Higgs couplings can be precisely quantified in supersymmetric models. For the case of the MSSM, we consider two ways to raise the Higgs mass to $126 \GeV$: An additional $D$-term allows the quartic Higgs coupling to be sufficiently increased already at tree-level~\cite{Batra:2003nj, Bellazzini:2009ix, Medina:2009ey}. Stops can then remain as light as possible consistent with the latest collider bounds. Alternatively, loop corrections from the stop sector can raise the quartic coupling to the required value. This, however, comes with an increased fine-tuning due to heavy stops. The Higgs sector of the MSSM contains two $CP$-even states. As the heavier state is decoupled, the Higgs couplings become more SM-like. We calculate the fine-tuning measure as a function of the mass of this state and find that it increases quadratically with the mass. Interestingly, however, this increase in fine-tuning can be offset with large $\tan\beta$. In the limit of large $\tan\beta$, the fine-tuning is therefore dominated by the usual contribution from stops, gluinos and Higgsinos (as well as any contribution from the additional $D$-term sector). Thus there is not necessarily any additional fine-tuning from having a Higgs with SM-like couplings in the MSSM.

This is no longer the case in the NMSSM which has an additional singlet superfield. This singlet gives a contribution to the quartic coupling of the Higgs and thereby allows its mass to already be raised at tree-level. In order to raise the mass to $126 \GeV$, a relatively large Higgs-singlet coupling near one is required~\cite{Barbieri:2006bg, Franceschini:2010qz, Hall:2011aa}. The Higgs sector now consists of three $CP$-even states.
The limit of SM Higgs couplings is obtained by decoupling the two heavy states, one of which is the singlet. We consider two versions of the NMSSM, 
one with a superpotential that has explicit mass terms and the other with a scale-invariant superpotential. Since the latter has no dimensionful parameters at the renormalizable level, it has the advantage of addressing the $\mu$-problem.
We derive the leading contribution to the fine-tuning measure from an expansion for large masses of the two heavy states.
We find that for both superpotentials the fine-tuning measure grows with the mass of the second $CP$-even state which is already present in the MSSM. It does not, on the other hand, grow with the singlet mass. 
For a Higgs-singlet coupling near one, constraints from electroweak precision tests (primarily due to the $T$-parameter) limit $\tan \beta$ to small values. This means that the increase in fine-tuning from the decoupling of the heavy $CP$-even states can no longer be compensated with large $\tan \beta$. We verify our approximations in deriving the leading contributions to the fine-tuning measure with a numerical scan over the parameter space for the scale-invariant superpotential. 

In the limit of weak mixing in the Higgs sector, we derive approximate formulas for the coupling of the lightest $CP$-even state (which we identify with the Higgs) to SM fermions and $W,Z$ gauge bosons. Using these formulas, we make a connection between the Higgs couplings and the fine-tuning measure and show that the tuning associated with the electroweak vev increases as the Higgs couplings become more SM-like. In particular we derive a lower bound on the tuning as a function of the deviations in the Higgs couplings from SM values. The achievable precision in the measurement of the Higgs couplings at the LHC and possibly the ILC was estimated in \cite{Peskin:2013xra}. We use these estimates to see what the LHC and ILC can teach us about the naturalness of the scale-invariant NMSSM. We find that the LHC may probe the naturalness of this model down to the few-percent level. For the model considered in \cite{Gherghetta:2012gb} with a low messenger scale ($20 \TeV$) and split families, this is comparable to the level of fine-tuning that can be deduced from direct searches at the LHC. The ILC, on the other hand, may probe naturalness down to the per-mille level. For this collider, Higgs coupling measurements can become the primary means of constraining the naturalness of supersymmetric models.

The results obtained in this paper complement previous work in ref.~\cite{Farina:2013fsa} which also found that Higgs coupling measurements will test the naturalness of the NMSSM with a large Higgs-singlet coupling. An analysis of Higgs couplings in supersymmetric models has also been done in~\cite{Gupta:2012mi,Blum:2012ii,D'Agnolo:2012mj,Gupta:2012fy,Barbieri:2013hxa}.

The paper is organised as follows. In \Secref{sec:MSSM}, we analyse the fine-tuning in the MSSM for two scenarios, first including additional $D$-terms to raise the Higgs mass and then with large loop corrections from the stop sector. The fine-tuning in the NMSSM is discussed in \Secref{sectionNMSSM}. We consider two different superpotentials, with and without explicit mass terms. For the scale-invariant superpotential, we derive a lower bound on the fine-tuning and analyse the implications for naturalness of future measurements of the Higgs couplings. In \Secref{sec:conclusions}, we present our conclusions. Technical details of the calculation are summarised in the appendix.

\section{The MSSM}
\label{sec:MSSM}
We consider the MSSM with an explicit $\mu$-term in the superpotential,
\be \label{mu-term}
W \, \supset \, \mu \, H_u H_d \, ,
\ee
where $H_u$ and $H_d$ are the two Higgs superfields. Including soft terms and the $D$-term contribution, 
the potential for the electromagnetically neutral Higgs scalars $H_u^0$ and $H_d^0$ is then given by
\be
V \, = \, (|\mu|^2 +m_{H_u}^2)  \, |H_u^0|^2 \, + \, (|\mu|^2 +m_{H_d}^2) \, |H_d^0|^2  \, - \, (B_\mu \, H_u^0 H_d^0 + \text{h.c.}) \, + \, \tilde{g}^2 \, (|H_u^0|^2 - |H_d^0|^2)^2 \, .
\label{HiggsPotential}
\ee
Using field redefinitions, $B_\mu$ and the vevs $\smash{\langle H^0_{u} \rangle}$ and $\smash{\langle H^0_{d} \rangle}$ can be chosen real and positive.

As is well known, the quartic coupling in the MSSM ($\tilde{g}^2 =  (g_1^2 + g_2^2)/8$, where $g_1$ and $g_2$ are the gauge couplings of, respectively, U(1)$_Y$ and SU(2)$_L$) is too small to give a Higgs mass of $126 \GeV$. An additional contribution is thus required. We will first discuss additional $D$-terms to raise the Higgs mass in \Secref{D-term-section} and then consider stop-loop corrections in \Secref{stop-loops-section}.

\subsection{Using $D$-terms to raise the quartic coupling}
\label{D-term-section}
\subsubsection{The Higgs sector}

In this section, we shall assume additional $D$-terms in order to lift the quartic coupling to the required value \cite{Batra:2003nj}. For example, assume a two-site moose model with gauge group $\text{SU}(N)_A  \times  \text{SU}(N)_B$, where the Higgs doublets $H_u$ and $H_d$ form a vector representation under SU$(N)_A$ (so that the $\mu$-term \eqref{mu-term} is allowed). Two link fields $\Sigma$ and $\tilde{\Sigma}$ in the bi-fundamental representation break this group down to the electroweak group SU$(2)_L$. 
The quartic coupling $\tilde{g}^2$ is then given by (see for example, \cite{Blum:2012ii})
\be
\tilde{g}^2 \, = \, \frac{1}{8} \left( g_1^2 \, (1 + \Delta_1) + g_2^2 \, (1 + \Delta_2) \right) \, ,
\ee
where $\Delta_1$ and $\Delta_2$ parameterize the contribution from the additional $D$-terms and are determined by the gauge couplings, gaugino masses and the breaking scale of the $\text{SU}(N)_A  \times  \text{SU}(N)_B$-sector. For our purposes, it is enough to know that realistic models can achieve $\Delta_1$ and $\Delta_2$ sufficiently large to raise the quartic coupling to the required value (see \cite{Batra:2003nj,Blum:2012ii} for more details). In this section, we will therefore treat $\tilde{g}$ as a free parameter whose value is fixed by the requirement that $m_{\rm Higgs} \approx 126 \GeV$. Since loop corrections, e.g.~from the stop sector, are then not required to raise $\tilde{g}^2$, we shall assume that their effect on the potential is small and will here correspondingly work with the tree-level potential~\eqref{HiggsPotential}. Corrections from the Coleman-Weinberg potential will be considered in sec.~\ref{finetuningstoploops}. In addition we assume that the underlying physics responsible for the $D$-term corrections does not increase the tuning.

We are interested in the limit where the couplings of the lightest $CP$-even Higgs are very similar to those of the SM Higgs.
To study this limit, it is convenient to rotate into the basis
\be
\label{basischange1a}
\begin{pmatrix}
\frac{h+i h_{I}}{\sqrt{2}}  \\ \frac{H+i H_I}{\sqrt{2}}
\end{pmatrix}
\, = \,
\begin{pmatrix}
\sin \beta & \cos \beta \\ -\cos \beta & \sin \beta  
\end{pmatrix}
 \begin{pmatrix}
H_u^0  \\ H_d^0 
\end{pmatrix} \, ,
\ee
where $\smash{\tan\beta \equiv \langle H^0_{u} \rangle/\langle H^0_{d} \rangle}$ and we have decomposed the fields into $CP$-even and $CP$-odd states.
The new basis is chosen such that $H$ does not obtain a vev, $\langle H \rangle=0$, and solely the vev of $h$ is responsible for electroweak symmetry breaking, $\smash{v \equiv \langle h \rangle = \sqrt{2} (\langle H^0_{u} \rangle^2 + \langle H^0_{d} \rangle^2)^{1/2} \simeq 246 \GeV}$. This state therefore couples to SM particles precisely like the SM Higgs. The Higgs thus becomes more like the SM Higgs, the larger its component of $h$. The admixture of the orthogonal state $H$, on the other hand, drives the couplings away from those in the SM.

We parameterize the mass matrix for the $CP$-even states $h$ and $H$ by
\be
\mathcal{M}^2=\begin{pmatrix}
m_h^2 & m_{hH}^2 \\ m_{hH}^2 &  m_H^2
\end{pmatrix} \, .
\ee
Expressions for the matrix elements are given in eq.~\eqref{potentialbasis2b} in appendix \ref{CPeveneigensystem}.
In the limit of small mixing, the mass of the lightest $CP$-even Higgs is well approximated by (see appendix~\ref{CPeveneigensystem})
\be \label{HiggsMassMSSM}
m_{\rm Higgs}^2 \, \simeq \,  m_h^2 - \frac{m_{hH}^4}{m_{H}^2}  \, . 
\ee
We can decouple $h$ from $H$ (and thus make the Higgs more SM-like) either by making the off-diagonal matrix element $m_{hH}^2$ smaller or by making the diagonal element $m_{H}^2$ larger. We define
\be \label{CouplingRatioDef}
r_u \, \equiv \, \frac{\text{Higgs coupling to up-type fermions}}{\text{SM Higgs coupling to up-type fermions}}
\ee
and similarly the coupling ratios $r_d$ to down-type fermions and $r_V$ to SM gauge bosons. To order $m_{hH}^4/m_H^4$, we have (see appendix \ref{app:HiggsCouplings}; see also \cite{Gupta:2012fy}):
\begin{subequations} \label{rurdrvMSSM}
\begin{align} 
& r_u \, \simeq \, 1 + \cot \beta \, \left(\frac{m_{hH}^2}{m_H^2}+\frac{m_h^2 m_{hH}^2}{m_H^4}\right)   - \frac{m_{hH}^4}{2 m_H^4}  \label{ru} \\
& r_d \, \simeq \, 1 - \tan \beta \, \left(\frac{m_{hH}^2}{m_H^2}+\frac{m_h^2 m_{hH}^2}{m_H^4}\right)  - \frac{m_{hH}^4}{2 m_H^4}  \label{rd}  \\
& r_V \, \simeq \, 1 - \frac{m_{hH}^4}{2 m_H^4}  \label{rv} \, .
\end{align}
\end{subequations}
We see that the Higgs couplings become indeed like in the SM in the limit $m_{hH}^2/m_H^2\rightarrow 0$. 

\subsubsection{The fine-tuning measure}
We seek to quantify whether taking the limit of SM Higgs couplings requires an increase in the fine-tuning. To this end, we define the fine-tuning measure as \cite{Barbieri:1987fn}
\be
\Sigma \, \equiv \, \sqrt{\sum_{\xi} \left( \frac{d \log v^2}{d \log \xi} \right)^2} \,, \label{finetuningdef}
\ee
where the sum runs over the parameters $\xi \in \{m_{H_u}^2, m_{H_d}^2, \mu, B_\mu, \tilde{g}\}$ that determine the Higgs potential eq.~\eqref{HiggsPotential}. We are mainly interested in the connection between naturalness and Higgs couplings which are in turn determined by the Higgs-sector parameters at low scales. Thus we choose these parameters to be evaluated at some low scale, like the electroweak scale.\footnote{An analogous definition for the fine-tuning measure, where all quantities are evaluated at the electroweak scale, has recently been considered in \cite{Baer:2012up}. We emphasise, however, that the effects from RG running can increase the fine-tuning significantly (e.g.~the factor in eq.~\eqref{UVrunningeffect} can be large).} Our definition of the fine-tuning measure therefore does not take any loop effects from the RG running above the electroweak scale, e.g.~due to heavy stops, into account. We will comment on the effect of the RG running on the fine-tuning in sec.~\ref{UVrunningsection}.

In order to evaluate the fine-tuning measure \eqref{finetuningdef}, we need to calculate the logarithmic derivatives of the Higgs vev $v$ with respect to the input parameters $\xi$. These steps are given in appendix~\ref{Finetuningexpressions}. 
We find
\begin{equation}\label{finetuning2}
\Sigma \, = \,  \sqrt{\sum_\xi \left(\frac{2\xi}{v}\frac{d v}{d \xi}\right)^2} = \frac{2}{v \det \mathcal{M}^2 }\sqrt{\sum_\xi \left(\xi \left[m_{hH}^2\ell_H' -m_{H}^2 \ell_h' \right]\right)^2} \, ,
\end{equation}
where $\det \mathcal{M}^2=m_h^2 m_H^2-m_{hH}^4$, $\ell_h\equiv\left.\partial V/\partial h\right|_\mathrm{min}$ and similarly for $\ell_H$ (see eq.~\eqref{minconditionsMSSM} for the explicit expressions) and $\ell'_{h,H} \equiv \partial \ell_{h,H} / \partial \xi$. The minimisation conditions for the Higgs potential are given by $\ell_{h,H}=0$.
This expression for the fine-tuning measure is general and loop corrections can be easily included by using the loop-corrected masses and derivatives of the loop-corrected potential.

Notice that part of the expression in eq.~\eqref{finetuning2} is given in terms of the mass matrix elements. If we manage to also express the remaining pieces, namely $\ell'_{h,H}$ and $\xi$, in terms of these masses, we can make a connection to the Higgs couplings via eq.~\eqref{rurdrvMSSM}. The Higgs potential \eqref{HiggsPotential} is determined by the parameters 
$ \{ m_{H_u}^2, m_{H_d}^2, \mu, B_\mu, \tilde{g} \}$.
In order to express the fine-tuning measure in terms of the mass matrix elements, we shall transform from this set of input parameters to a new set of input parameters. First, we can use the minimisation conditions for the potential (see eq.~\eqref{minconditionsMSSM}) to express $m_{H_u}^2$ and $m_{H_d}^2$ in terms of $v$ and $\tan \beta$ (and $\mu, B_\mu, \tilde{g}$). We can thus consider $ \{ v, \tan \beta, \mu, B_\mu, \tilde{g} \}$ as a basis of input parameters. In terms of these parameters, the elements of the mass matrix are given by
\begin{subequations}\label{MassMatrixElementsDtermMSSM}
\begin{align}
& m_h^2 \, = \, 2 \, \tilde{g}^2 v^2 \cos^2 2 \beta \label{Higgsmassbasis1}\\ 
& m_H^2 \, = \, 2 \, \tilde{g}^2 v^2 \sin^2 \hspace*{-0.035cm} 2 \beta+ 2 \, B_\mu \csc 2 \beta \label{Hmassbasis1} \\
& m_{hH}^2 \, = \, \tilde{g}^2 v^2 \sin 4 \beta\, \label{mixingmassbasis1}.
\end{align}
\end{subequations}

It will be convenient to express the coupling $\tilde{g}$ in terms of the mass matrix element $m_h^2$ associated with the $CP$-even state $h$ (which becomes the Higgs mass in the limit of vanishing mixing, see eq.~\eqref{HiggsMassMSSM}). Solving eq.~\eqref{Higgsmassbasis1} for $\tilde{g}$, we find
\be
\label{grelation}
\tilde{g} \, = \, \frac{m_h}{\sqrt{2} \,  v |\cos 2 \beta|} \, .
\ee
Similarly we can use eq.~\eqref{Hmassbasis1} to fix the soft mass parameter $B_\mu$ in terms of the mass matrix element $m_H^2$ associated with the $CP$-even state $H$:
\be
\label{brelation}
B_\mu \, = \, \frac{m_H^2 - m_h^2  \,\tan^2 \hspace*{-0.035cm} 2 \beta }{2  \csc 2 \beta } \, .
\ee
With the help of these two relations, we can express all quantities in terms of the parameters
$ \{ v, \tan \beta, \mu, m_h, m_H \}$.
In particular we find
\be
m_{hH}^2 \, = \,  m_h^2 \,  \tan 2 \beta  \, = \frac{2\tan\beta}{1-\tan^2 \hspace*{-0.035cm}  \beta} \, m_h^2 .
\label{mixingmassterm}
\ee

Applying eqs.~\eqref{grelation} and \eqref{brelation} to $\ell'_{h,H}$ and $\xi$ in eq.~\eqref{finetuning2}, we obtain an analytic expression for the fine-tuning measure as a function of these new input parameters. This expression is given in eq.~\eqref{SigmaMSSM} in appendix \ref{Finetuningexpressions}. 
We are interested in the limit of a SM Higgs, corresponding to $m_H^2 \gg |m_{hH}^2|$ according to~eq.~\eqref{rurdrvMSSM}. Let us first consider the case where $\tan \beta = \mathcal{O}(1)$. From eq.~\eqref{mixingmassterm}, we see that then $|m_{hH}^2| \sim m_h^2$ and the decoupling limit thus requires that $m_H \gg  m_h$. Expanding the fine-tuning measure for large $m_H$, the leading term in $m_H$ is given by
\be
\Sigma \, \approx \, \sqrt{\frac{3}{2}} \, \sin^2 \hspace*{-0.035cm} 2 \beta \, \frac{m_H^2}{m_h^2} \, + \, \mathcal{O}(m_H^0) \, .
\label{fine-tuning-limit}
\ee
We see that
the fine-tuning grows like $m_H^2$. This can be understood from the fact that, for $\smash{\tan \beta = \mathcal{O}(1)}$, $m_H \gg m_h$ in turn requires that $B_\mu \gg m_h^2$ (see eq.~\eqref{Hmassbasis1}).
This large soft mass parameter results in a large fine-tuning.

In the opposite case $\tan \beta \gg 1$, the contribution \eqref{fine-tuning-limit} to the fine-tuning measure is suppressed with $\smash{\sin^2 \hspace*{-0.035cm} 2 \beta \approx 4/\tan^2 \hspace*{-0.035cm} \beta }$. This is related to the fact that $\smash{m_H^2 \approx B_\mu \tan \beta}$ for large $\tan \beta$. A given $m_H$ in this case thus corresponds to a smaller $B_\mu$ compared to the case $\tan \beta = \mathcal{O}(1)$. Accordingly the fine-tuning is smaller. Note that this argument assumes, however, that large $\tan \beta$ does not itself increase the fine-tuning. It is straightforward to see from the Higgs potential that it indeed does not: The relevant part of the potential involving the field that obtains a vev, $h$, can be written as $\smash{V \supset m^2 h^2 + \sigma h^4}$. In the limit $\tan \beta \gg 1$, we find at leading order that $m^2 \approx (m_{H_u}^2 + \mu^2)/2$ and $\sigma \approx \tilde{g}^2 /4$. Any fine-tuning that is necessary in order to obtain the correct Higgs vev arises between the two terms in the expression for $m^2$. The fact that neither of these terms is enhanced with $\tan \beta$ shows that the fine-tuning does not grow with $\tan \beta$ either. 
Expanding the fine-tuning measure for large $\tan \beta$ (instead of for large $m_H$), we find
\be 
\Sigma \, \approx \, \sqrt{\frac{20 \mu^4}{m_h^4}  + \frac{4 \mu^2}{m_h^2} + 5 } \, + \, \mathcal{O}(\frac{1}{\tan^2 \hspace*{-0.035cm}  \beta}) \, . 
\label{fine-tuning-large-tanbeta}
\ee
In this limit, the fine-tuning is thus dominated by the $\mu$-term which is unrelated to the Higgs couplings.

In addition, the mixing mass $m_{hH}^2$ is suppressed for large $\tan \beta$ since $m_{hH}^2 \approx - 2 m_h^2 / \tan \beta$. The decoupling limit $m_H^2 \gg |m_{hH}^2|$ can thus be achieved even for $m_H \sim m_h$ if $\tan \beta$ is large. 
This is an alternative direction in the parameter space $ \{ v, \tan \beta, \mu, m_h, m_H \}$ along which the Higgs couplings become more SM-like (but the fine-tuning is not increased).

Let us finally combine the approximation \eqref{fine-tuning-limit} for the fine-tuning measure with the approximations \eqref{rurdrvMSSM} for the coupling ratios. To this end, note that we can neglect the terms of order $m_H^{-4}$ in \eqref{ru} and \eqref{rd} even for large $\tan \beta$ because they are suppressed by at least an additional factor of $m_h^2 / m_H^2$ compared to the term of order $m_H^{-2}$.
Solving the resulting simpler relations for $m_H^2$, we can express the fine-tuning measure in terms of either $r_u$, $r_d$ or $r_V$:\footnote{We thus make another transformation from $m_H^2$ to $r_u$ and thus to the basis $ \{ v, \tan \beta, \mu, m_h, r_u\}$ of input parameters and similarly for $r_d$ and $r_V$.}
\be 
\Sigma  \, \approx \,  
\begin{cases}\label{ftapproxMSSM} 
  \frac{4 \sqrt{6}}{(1-r_u) \, \tan^4 \hspace*{-0.035cm} \beta }  \\
\frac{4 \sqrt{6}}{(r_d-1) \, \tan^2 \hspace*{-0.035cm} \beta }  \\
\frac{4 \sqrt{3}}{\sqrt{1-r_V} \tan^3 \hspace*{-0.035cm} \beta} \, .
\end{cases}
\ee
We have only kept the leading $\tan \beta$-dependence in the leading terms in $r_{u,d,V}-1$ (coming from eq.~\eqref{fine-tuning-limit}). For very large $\tan \beta$, eventually \eqref{fine-tuning-large-tanbeta} will dominate the fine-tuning measure. 
Note that for $m_H \gg v$, the coupling ratios satisfy $r_{u,V}< 1$ and $r_d > 1$ as follows from eq.~\eqref{rurdrvMSSM}. This ensures that \Eqref{ftapproxMSSM} is positive.
We now see explicitly that for fixed $\tan \beta$, the fine-tuning grows in the SM limit, $r_{u,d,V}\rightarrow 1$. As we have discussed, however, large $\tan \beta$ is an alternative direction in the parameter space $ \{v, \tan \beta, \mu, m_h, m_H\}$ along which the Higgs becomes more SM-like but the fine-tuning does not increase. In eq.~\eqref{ftapproxMSSM}, this is manifest in the $\tan \beta$-dependent suppression. We thus conclude that SM-like couplings do not necessarily imply larger fine-tuning in the MSSM.

In sec.~\ref{sectionNMSSM}, we shall investigate the NMSSM, where large $\tan \beta$ is typically in conflict with electroweak precision tests when the Higgs-singlet coupling $\lambda$ is used to raise the Higgs mass to the required value.

\subsection{Using stop-loop corrections to raise the quartic coupling}
\label{stop-loops-section}
\subsubsection{The fine-tuning measure}
\label{finetuningstoploops}

We shall now consider the case in which the quartic coupling in the Higgs potential is raised via loop corrections from the stop sector (instead of additional $D$-terms). The Higgs potential at tree-level is given by \Eqref{HiggsPotential} with $\tilde{g}^2 = (g_1^2 + g_2^2)/8$. We will only consider the dominant loop corrections coming from the top/stop sector. At one-loop order, the correction to the tree-level potential is given by the Coleman-Weinberg potential
\be \label{CW}
V_{\rm CW} \, = \,  \, \sum_{i=1,2} \frac{3 \, m_{\tilde{t}_i}^4}{32 \pi^2}  \left(\log \frac{m_{\tilde{t}_i}^2}{\mu_r^2} - \frac{3}{2} \right)- \frac{6\, m_t^4}{32 \pi^2} \left(\log \frac{m_t^2}{\mu_r^2} - \frac{3}{2} \right) \, ,
\ee
where $m_{\tilde{t}_{1,2}}$ denote the two stop masses and $m_t$ is the top mass. The Coleman-Weinberg potential depends on the Higgs vev via these masses. The stop masses are also determined by the soft parameters of the stop sector, $\smash{m_{Q_3}^2}$, $\smash{m_{u_3}^2}$ and $A_t$. All parameters in the potential are evaluated at the renormalization scale $\mu_r$ in the $\smash{\overline{DR}}$-scheme. In order to minimise the logarithms, we will choose $\mu_r=\sqrt{ m_{Q_3} m_{u_3}}$.

The Higgs potential is now determined by $\smash{\{m_{H_u}^2, m_{H_d}^2, \mu, B_\mu, \tilde{g}, m_{Q_3}^2, m_{u_3}^2, A_t\}}$. These parameters constitute the set $\xi$ over which we sum when evaluating the fine-tuning measure $\eqref{finetuningdef}$.
The expression for the derivatives $d v / d \xi$ that we derive in appendix \ref{Finetuningexpressions} remains applicable when we include the Coleman-Weinberg potential. We can thus use eq.~\eqref{finetuning2} to evaluate the fine-tuning measure also for this case. There are two differences compared to the tree-level case: The mass matrix elements that appear in \eqref{finetuning2} now include the loop corrections from the Coleman-Weinberg potential. In addition the expressions for $\ell_{h,H}$ are modified to $\smash{\ell_h = \ell_h^{\rm tree} + \partial V_{\rm CW} / \partial h}$ and similarly for $H$, where $\ell_{h,H}^{\rm tree}$ are the corresponding tree-level expressions given in eq.~\eqref{minconditionsMSSM}. Let us emphasise that this works more generally: Taking the aforementioned changes into account, eq.~\eqref{finetuning2} allows us to find an analytical expression for the fine-tuning measure for any effective potential (not just that in eq.~\eqref{CW}; see e.g.~\cite{Cassel:2010px,Baer:2012up} for earlier, related work).

In order to make a connection with the Higgs couplings, we want to also express $\ell^\prime_{h,H}$ and $\xi$ (which appear in eq.~\eqref{finetuning2}) in terms of Higgs sector masses. To this end, we will again transform to a new basis of input parameters. 
Using the minimisation conditions $\ell_{h,H}= 0$, we first trade $m_{H_u}^2$ and $m_{H_d}^2$ for $v$ and $\tan \beta$. Expressions in this basis for the mass matrix elements including the loop corrections from the top/stop sector can be found in \cite{Carena:1995bx}. In order to maximise the tree-level contribution to the Higgs mass (cf.~eq.~\eqref{Higgsmassbasis1}), we will assume that $\smash{\tan\beta\gg 1}$. For simplicity we will only report terms up to order $(1/\tan \beta)^0$:
\begin{subequations}
\begin{align}
& m^2_h  \, \simeq \, 2 \tilde{g}^2 v^2 + \frac{3}{8\pi^2} y_t^2 (y_t^2-2\tilde{g}^2)  v^2\log\frac{\mu^2_r}{m^2_t}+\frac{3}{16\pi^2} y_t^4 v^2 X_t \label{stoploopmhsq} \\
& m^2_H  \simeq \,  B_{\mu} \, \tan\beta -\frac{y_t^4v^2 A_t^2  \mu^2}{32\pi^2 \mu^4_r}
 \label{stoploopmHsq} \\
& \label{stoploopmhHsq}  m^2_{hH}   \, \simeq \,  \frac{y_t^4 v^2\mu A_t (A_t^2-6 \mu_r^2)}{32 \pi^2 \mu^4_r} \, ,
\end{align}
\end{subequations}
where 
\begin{equation}
X_t\, \equiv \, \frac{2A_t^2}{\mu^2_r}\left(1-\frac{A_t^2}{12 \mu^2_r}\right) \, . 
\end{equation}
We next invert the relations for $m_h$ and $m_H$ in order to trade $\tilde{g}$ and $B_\mu$ for these masses. The resulting relations up to terms of order $1/ \tan \beta$ are:
\begin{subequations}
\begin{align}
\tilde{g}^2 & \, \simeq \,  \frac{1}{\left(1-\frac{3}{8\pi^2} y_t^2\log\frac{\mu_r^2}{m^2_t}\right)}\left[ \frac{m^2_h}{2v^2}-\frac{3 y_t^4}{16\pi^2}\left(\frac{X_t}{2}+\log\frac{\mu_r^2}{m^2_t}\right)
 - \frac{y_t^4 \mu A_t(A_t^2-6 \mu_r^2)}{16\pi^2 \mu_r^4 \tan\beta} 
\right]\\
 B_{\mu} & \, \simeq \,  \frac{1}{\tan\beta}\left[m^2_H+\frac{y_t^4\mu^2  v^2 A_t^2}{32 \pi^2 \mu_r^4}\right] \, .
 \label{modgbmu}
\end{align}
\end{subequations}

Using the relations for $\tilde{g}$ and $B_\mu$ (before expanding in $\tan \beta$) in eq.~\eqref{finetuning2}, we obtain the fine-tuning measure expressed in terms of the parameters $\smash{\{ v, \tan\beta, \mu, m_h, m_H, m^2_{Q_3},m^2_{u_3}, A_t \}}$. For simplicity, let us focus on the contribution of the soft mass $m_{H_u}^2$ to the fine-tuning measure. As we discuss in the next section, this contribution is enhanced when we take the loop corrections from stops during the RG running into account. Expanding in both $m^2_H$ and $\tan \beta$ and keeping only the leading terms in either $m^2_H$ or $\tan \beta$, we find
\begin{eqnarray}
\Sigma_{m^2_{H_u}} \equiv \left| \frac{d \log v^2}{d \log m_{H_u}^2} \right|  \, \approx \, \left| 
\frac{2 m^2_H}{m^2_h\tan^2 \hspace*{-0.035cm} \beta} -\frac{2 \tilde{\mu}^2}{m^2_h}\right| \, ,
\label{tottuningMSSM}
\end{eqnarray}
where
\be
\tilde{\mu}^2 \equiv  \mu^2 +  \frac{3\, v^2 y_t^4}{16\pi^2} \left[ 2 + 2 \log\frac{\mu_r^2}{m^2_t}
-\sum_{i=1,2} \frac{m^2_{\tilde{t}_i}}{m^2_{t}}\left(1-\log\frac{m^2_{\tilde{t}_i}}{\mu_r^2}\right) \left(1+(-1)^i \frac{A^2_t}{\sqrt{(m^2_{Q_3}-m^2_{u_3})^2+4 A^2_t v^2 y_t^2}} \right)
\right] \, .
\ee
Comparing with eqs.~\eqref{fine-tuning-limit} and \eqref{fine-tuning-large-tanbeta}, we see that we recover a similar behaviour of the fine-tuning measure as in the last section. In particular the leading dependence on both $m_H$ and $\mu$ remains unchanged. At leading order, the Coleman-Weinberg potential only affects the `effective $\mu$-term' $\tilde{\mu}$ which is relevant for the fine-tuning measure. 

An important difference compared to the case of negligible loop corrections, however, is the mixing mass $m_{hH}^2$. As we see from eq.~\eqref{stoploopmhHsq}, it no longer vanishes in the limit of large $\tan \beta$ if stop loops are important. Accordingly the limit of SM Higgs couplings necessarily requires large $m_H$. When we combine the loop-corrected fine-tuning measure (i.e.~\eqref{tottuningMSSM} and the other derivatives) with the relations for the coupling ratios \eqref{rurdrvMSSM}, this leads to a similar relation as \Eqref{ftapproxMSSM} but with one power of $\tan \beta$ less in the denominators. Nevertheless, even in this case we can compensate the increase in fine-tuning for $r_{u,d,V}\rightarrow 1$ with large $\tan \beta$.

\subsubsection{The effect of the RG running on the fine-tuning measure}
\label{UVrunningsection}

As we have discussed below eq.~\eqref{finetuningdef}, we evaluate the fine-tuning measure using derivatives with respect to parameters defined at some low scale (like the electroweak scale or, in this section, $\mu_r$). We thus do not include the effect on the fine-tuning of the RG running from some high scale (like the messenger scale) to that low scale. Since these corrections are of course important in the case of heavy stops, we will now discuss how they affect our estimate of the fine-tuning. To this end, let us use a different (`more standard') definition of the fine-tuning measure, where the derivatives are taken with respect to parameters $\widehat{\xi}$ defined at the messenger scale $\Lambda_{\rm mess}$. 
We thus replace $d \log v / d \log \xi$ in \eqref{finetuningdef} by $d \log v / d \log \widehat{\xi}$. In the following, a hat shall denote soft mass parameters defined at $\Lambda_{\rm mess}$, whereas un-hatted parameters are defined at $\mu_r$. 

The stops affect the Higgs sector during the RG running dominantly via the soft mass $m_{H_u}^2$. 
Neglecting loop contributions from other particles, we find at leading-log order that
\begin{equation}
m^{2}_{H_u} \, \simeq \, \widehat{m}^2_{H_u} - \Delta_{\tilde{t}} \qquad \text{where} \quad \Delta_{\tilde{t}} \, \equiv \,  \frac{3y_t^2}{8\pi^2}\left(\widehat{m}^2_{Q_3}+\widehat{m}^2_{u_3}+\widehat{A}_t^2\right)\log\frac{\Lambda_{\rm mess}}{\mu_r} \, .
\end{equation}
This correction contributes to the fine-tuning measure via the derivative with respect to the soft mass $\widehat{m}^2_{H_u}$.\footnote{The derivatives with respect to the soft mass parameters $\widehat{m}^2_{Q_3}$, $\widehat{m}^2_{u_3}$ and $\widehat{A}_t$ of the stop sector give a contribution of a similar size.} We find
\be
\Sigma^{\rm mess}_{m^2_{H_u}} \, \equiv \, \left|\frac{d \log v^2}{d \log \widehat{m}_{H_u}^{2}}\right| \, \approx \, \left| \frac{\widehat{m}_{H_u}^{2}}{m_{H_u}^{2}} \, \frac{d \log v^2}{d \log m_{H_u}^{2}} \frac{d \, m_{H_u}^{2}}{d \, \widehat{m}_{H_u}^{2}}\right| \, ,
\ee
where we have neglected additional, small terms arising according to the chain rule.\footnote{These terms are given by $(d \log v^2 / d \log \xi) (d \log \xi / d \log \widehat{m}_{H_u}^2)$ with $\xi \in \{m_{H_d}^2 ,B_\mu, \mu, \tilde{g}, m_{Q_3}^2, m_{u_3}^2, A_t\}$. For all these terms, either the first or the second derivative is small.} 
Together with the definition of $\Sigma_{m^2_{H_u}}$ in \eqref{tottuningMSSM}, this gives
\be \label{UVrunningeffect}
\Sigma^{\rm mess}_{m^2_{H_u}} \, \approx \, \left|1 + \frac{\Delta_{\tilde{t}}}{m_{H_u}^2} \right| \, \Sigma_{m^2_{H_u}} \, \sim \, \left|1 + \frac{ \Delta_{\tilde{t}} \, \tan^2 \hspace{-0.035cm}\beta }{\, m_H^2} \right| \, \Sigma_{m^2_{H_u}}  \, .
\ee
We see that the RG running increases the fine-tuning compared to our measure \eqref{finetuningdef} if $\Delta_{\tilde{t}} \gg m_{H_u}^2$. In the last step, we have expressed $m^2_{H_u}$ in terms of the basis $\smash{\{v, \tan \beta, \mu, m_h, m_H ,  m_{Q_3}^2, m_{u_3}^2, A_t\}}$ and expanded for large $m_H$ and $\tan \beta$. We see that for $m_H^2 \gg \Delta_{\tilde{t}}  \tan^2 \hspace{-0.035cm}\beta$, on the other hand, the effect of the RG running on the fine-tuning becomes negligible. This can be understood as follows: We can, roughly, distinguish two types of tuning that lead to the correct electroweak scale. In the first case, parameters at $\mu_r$ are much larger than the electroweak scale. In order to obtain the correct Higgs vev, the various parameters that enter the minimisation conditions for the Higgs potential need then to be tuned against each other. This happens e.g.~in the decoupling limit $m_H \gg m_h$ for small $\tan \beta$ since then also $B_\mu \gg m_h^2$ (cf.~eq.~\eqref{brelation}). In the second case, on the other hand, parameters at $\mu_r$ are of order the electroweak scale and no tuning in the minimisation conditions is required. If loop corrections to these parameters are large, however, a tuning among the various contributions that affect the RG evolution is necessary in order to make them small at $\mu_r$. Such large loop corrections arise e.g.~from heavy stops. Our fine-tuning measure \eqref{finetuningdef} captures only the first type of tuning whereas the additional factor in \eqref{UVrunningeffect} accounts for the second type. However, we see that when the soft parameters, e.g.~$B_\mu$,  are large at $\mu_r$, the RG running has less of an effect on the tuning. This explains the suppression of the correction to the factor in eq.~\eqref{UVrunningeffect} for large $m_H$. Alternatively, if the corrections from RG running are important and this factor is large, our measure \eqref{finetuningdef} provides a lower bound on the fine-tuning.

\section{The NMSSM}
\label{sectionNMSSM}

\subsection{The NMSSM with superpotential mass terms}
\label{NMSSMspmt}
The NMSSM has a singlet superfield $S$ in addition to the particle content of the MSSM. This allows for several new terms in the superpotential. 
In this section, we shall consider a superpotential with an explicit $\mu$-term and a singlet mass term:
\be
W \, \supset \, \lambda \, S H_u H_d \, + \, \mu \, H_u H_d \, + \, \frac{1}{2} \, M \, S^2 \, . \label{spmt}
\ee
This superpotential was studied in the context of large coupling $\lambda$ in \cite{Barbieri:2006bg}. Including soft terms and the $D$-term contributions, the potential for the electromagnetically neutral Higgs scalars $H^0_u$ and $H^0_d$ and the singlet scalar $S$ is given by
\begin{multline} \label{potentialNMSSMspmt}
V =  (m_{H_{u}}^2 + |\mu + \lambda S|^2) |H^0_{u}|^2\, + \,(m_{H_{d}}^2 + |\mu + \lambda S|^2) |H^0_d|^2 \, + \,| M S  -\lambda \, H^0_{u} H^0_{d} |^2 +m_S^2 |S|^2 \, \\
\, +\,  \left[-a_{\lambda} \, S H^0_{u} H^0_{d} - B_\mu \, H^0_{u} H^0_{d} +\text{h.c.} \right]\, + \, \tilde{g}^2 \, (|H_u^0|^2 - |H_d^0|^2)^2 \, .
\end{multline}
To ensure $CP$-conserving vevs, we shall assume that the Higgs potential is $CP$-invariant. Using redefinitions, we can then choose $\lambda$ and the vevs $\langle H^0_{u,d} \rangle$ real and positive. All other parameters in the Higgs potential are also real but can have both signs. 

As we will see explicitly below (cf.~eqs.~\eqref{mhsqrelNMSSMdsp} and \eqref{HiggsMassNMSSM}), the Higgs-singlet coupling $\lambda$ gives a welcome contribution to the Higgs mass which allows to raise it to $126 \GeV$ already at tree-level. This requires that $\lambda \sim 1$ \cite{Barbieri:2006bg,Gherghetta:2012gb} which we assume in the following.\footnote{Note, however, that for such a large value at the electroweak scale, the coupling $\lambda$ hits a Landau pole before the GUT scale. Some UV completion has to kick in before this Landau pole. Possible UV completions where discussed e.g.~in \cite{Harnik:2003rs,Gherghetta:2011wc}.\label{lambda-value-footnote}} Since finite loop corrections from the stop sector are no longer required to raise the Higgs mass, we expect them to be small and neglect them. We similarly neglect finite loop corrections to the potential from the Higgs-singlet sector which can become important for large $\lambda$ and $m_{H_u}^2, m_{H_d}^2, m_S^2$. A proper treatment of these corrections is, however, quite involved and beyond the scope of this paper. Finally, we again use \Eqref{finetuningdef} as the definition for our fine-tuning measure. Accordingly, we do not account for loop corrections during the RG running. But similar to what we have discussed in sec.~\ref{UVrunningsection}, the effect of top/stop loops on the fine-tuning becomes less important, the larger the masses of the heavy Higgses are. Since this is ultimately the limit that we are interested in, we expect that these corrections will not significantly affect our fine-tuning estimates. Alternatively, if these corrections are important, the fine-tuning is larger than what we calculate using our measure \eqref{finetuningdef} (cf.~eq.~\eqref{UVrunningeffect}). In this case, our measure provides a lower bound on the fine-tuning. Loop corrections from the RG running of, for example, $m^2_{H_u}$ and $m^2_{H_d}$ proportional to $\lambda^2 m^2_S$ may not be suppressed for large heavy Higgs masses. However, as long as the messenger scale is well below the energy scale where $\lambda$ develops a Landau pole, these corrections are suppressed by a loop factor with respect to the tree-level result. 

Similar to our discussion for the MSSM, we rotate the fields into the basis
\be
\label{basischange1aNMSSM}
\begin{pmatrix}
\frac{h+i h_I}{\sqrt{2}}  \\ \frac{H+i H_I}{\sqrt{2}} \\ \frac{s+i s_I}{\sqrt{2}}
\end{pmatrix}
\, = \,
\begin{pmatrix}
\sin \beta & \cos \beta & 0 \\ -\cos \beta & \sin \beta  & 0 \\
0 & 0 & 1
\end{pmatrix}
 \begin{pmatrix}
H_u^0  \\ H^0_d \\ S
\end{pmatrix} \, ,
\ee
where $H$ does not obtain a vev. 
The $CP$-even state $h$ thus couples at tree-level precisely like the SM Higgs, whereas an admixture with $H$ and $s$ drives the couplings away from the SM limit. We shall denote the vev of the singlet scalar as $\smash{v_s \equiv \langle s \rangle = \sqrt{2} \langle S \rangle}$.

We parameterize the mass matrix for the $CP$-even Higgs states by
\be
\mathcal{M}^2= \begin{pmatrix}
m_h^2 & m_{hH}^2 & m_{hs}^2\\ 
m_{hH}^2 &  m_H^2 & m_{Hs}^2\\ 
m_{hs}^2 &  m_{Hs}^2 & m_{s}^2
\end{pmatrix} \, . \label{massmatrixNMSSM}
\ee
Expressions for the mass matrix elements are given in eq.~\eqref{MassMatrixElementsNMSSMdsp} in appendix~\ref{CPeveneigensystem}. 
We are again interested in the limit of a Higgs with SM couplings or, correspondingly, the limit of vanishing mixing of $h$ with $H$ and $s$. Fixing $m_h^2$, this limit is approached either by raising the diagonal matrix elements $m_H^2$ and $m_s^2$ or by lowering the off-diagonal elements $m_{hH}^2$ and $m _{hs}^2$. In the limit of small mixing, the Higgs mass is well approximated by (see appendix~\ref{CPeveneigensystem})
\be \label{HiggsMassNMSSM}
m_{\rm Higgs}^2 \, \simeq \,  m_h^2 - \frac{m_{hH}^4}{m_{H}^2} - \frac{m_{hs}^4}{m_{s}^2} \, . 
\ee

The Higgs potential eq.~\eqref{potentialNMSSMspmt} is determined by the parameters
$ \{ m_{H_u}^2, m_{H_d}^2, m_S^2, \lambda, a_\lambda, \mu, B_\mu, M, \tilde{g} \} $. These parameters constitute the set $\xi$ over which we sum when calculating the fine-tuning measure \eqref{finetuningdef}. 
In order to evaluate the fine-tuning measure, we need to determine the derivatives $d v / d  \xi$. This is described in appendix~\ref{Finetuningexpressions}. The resulting expression for $d v / d  \xi$ (see eq.~\eqref{dvdxiNMSSM}) is again partially given in terms of the mass matrix elements. In order to make the connection to the Higgs couplings, we want to express also the remaining pieces in terms of these masses. 
To this end, we shall again transform to a new set of input parameters. Using the minimisation conditions $\ell_{h,H,s}=0$ for the potential (see~eq.~\eqref{minconditionsNMSSMdsp} for the explicit expressions), we first obtain
$ \{ v, \tan \beta, v_s, \lambda, a_\lambda, \mu, B_\mu, M, \tilde{g} \} $
as an alternative parameter set. Expressed in terms of these parameters, the elements of the Higgs mass matrix read
\begin{subequations}
\begin{align}
& m_h^2 \, = \, 2 \, \tilde{g}^2 v^2 \cos^2 \hspace*{-0.035cm} 2 \beta \, + \, \frac{1}{2} \lambda^2 v^2 \sin^2 \hspace*{-0.035cm} 2 \beta \label{mhsqrelNMSSMdsp}  \\
& m_H^2 \, = \, \frac{1}{2}  \,\csc 2 \beta \,   \left( 2 \sqrt{2} \, v_s\, (a_\lambda + \lambda  M)+ 4 B_\mu  - v^2 \sin^3 \hspace*{-0.035cm} 2 \beta  \, \left(\lambda^2 - 4 \tilde{g}^2 \right)\right) \label{mHsqrelNMSSMdsp}  \\
& m_s^2 \, = \, \frac{v^2}{\sqrt{2} \, v_s} \, \bigl(\sin \beta \,  \cos \beta \, (a_\lambda + \lambda  M) - \lambda \, \mu \bigr)\label{mssqrelNMSSMdsp}  \\
& m_{h s}^2 \, = \, v  \left(   \lambda^2 v_s + \sqrt{2} \lambda  \mu  -  \sqrt{2} \, \sin \beta \, \cos \beta \, (a_\lambda +\lambda  M) \right)  \label{mshsqrelNMSSMdsp} \\
& m_{h H}^2 \, = \, \frac{v^2}{4} \sin 4 \beta \,  \left(4 \tilde{g}^2- \lambda^2 \right) \label{mhHsqrelNMSSMdsp}  \\
& m_{H s}^2 \, = \, \frac{v}{\sqrt{2}} \cos 2 \beta \,  (a_\lambda + \lambda  M) \label{msHsqrelNMSSMdsp}  \, . 
\end{align}
\end{subequations}

We shall again invert these relations in order to express some of the input parameters in terms of Higgs-sector masses. Notice that $a_\lambda$ and $M$ always appear in the combination $a_\lambda + \lambda M$. We can therefore solve for only one of these two parameters which we choose to be $a_\lambda$. Using the relations for $m_H^2$, $m_s^2$, $m_{hs}^2$ and $m_{Hs}^2$, we can in addition solve for $v_s$, $\mu$ and $B_\mu$. This gives
\begin{subequations}
\begin{align}
& v_s \, = \, \frac{m_{hs}^2 v}{ \lambda^2 v^2 - 2 m_s^2 } \\ 
& \mu \, = \,  \frac{2  m_s^2 \, m_{hs}^2+ m_{Hs}^2 \tan 2 \beta  \left(2 m_s^2 - \lambda^2 v^2\right)}{\sqrt{2} \lambda v\, (2 \, m_s^2 -  \, \lambda^2 v^2)} \\ 
& B_\mu \, = \, \frac{m_{hs}^2 m_{Hs}^2}{2 m_s^2- \lambda^2 v^2} \sec 2 \beta - \frac{1}{4} \sin 2 \beta \left( v^2 \left(4 \tilde{g}^2 - \lambda^2 \right)\sin^2 \hspace*{-0.035cm} 2 \beta  \, -2  m_H^2 \right)\\
& a_\lambda \, = \, \sqrt{2} \frac{m_{Hs}^2}{v} \sec 2 \beta - \lambda  M \, .
\end{align}
\end{subequations}
With the help of these relations, we can express all quantities in terms of the new basis of input parameters
$ \{ v,  \tan \beta, \lambda,  \tilde{g}, M, m_H, m_s, m_{hs}^2, m_{Hs}^2 \}. $

The expression for the fine-tuning measure in this basis is rather lengthy and will therefore not be reported here. We can approach the decoupling limit for $H$ and $s$, $\smash{m_H^2, m_s^2 \gg |m_{hH}^2|, |m_{hs}^2|}$, either with large diagonal matrix elements $m_H^2$ and $m_s^2$ or with small off-diagonal matrix elements $\smash{|m_{hH}^2|}$ and $\smash{|m_{hs}^2|}$. Let us first consider the former case. Assuming that also $\smash{m_H^2, m_s^2 \gg M^2}$, we can expand the fine-tuning measure in $m_H^2$ and $m_s^2$ and find
\be\label{fine-tuning-limit-NMSSMdsp}
\Sigma \, \approx \, \sqrt{\frac{3}{2}} \, \sin^2 \hspace*{-0.035cm} 2 \beta \, \frac{m_H^2}{m_h^2}  \, + \, \mathcal{O}\Bigl(\frac{m_H^2}{m_s^2},m_H^0\Bigr)\, .
\ee
Notice that the leading term is independent of $m_s$. An explicit dependence on this mass only arises at order $m_H^2/m_s^2$. 
This can be understood as follows: From eq.~\eqref{MassMatrixElementsNMSSMdsp}, we see that we can raise the mass $m_s$ while keeping the other mass matrix elements fixed by raising the soft mass $m_S^2$. In the limit of very large $m_S^2$, the singlet can be integrated out. One finds that the resulting potential is just the MSSM potential plus a $\lambda$-dependent contribution to the quartic coupling. This shows that even in the limit $m_s \rightarrow \infty$, the fine-tuning stays finite at tree-level.\footnote{Note, however, that a large soft mass $m_S^2$ feeds into $m_{H_u}^2$ and $m_{H_d}^2$ during the RG running and thereby increases the fine-tuning (see e.g.~\cite{Hall:2011aa}). 
\label{large-ms-footnote}} The dependence of our fine-tuning measure at tree-level on $m_s$ can accordingly only enter at order $\smash{m_s^{-|x|}}$. 

Furthermore, notice that the leading contribution to the fine-tuning measure, eq.~\eqref{fine-tuning-limit-NMSSMdsp}, has the same form as in the MSSM, eq.~\eqref{fine-tuning-limit}. The only difference arises from the different expressions for $m_h$ and $m_H$ (eqs.~\eqref{Higgsmassbasis1} and \eqref{mhsqrelNMSSMdsp} and eqs.~\eqref{Hmassbasis1} and \eqref{mHsqrelNMSSMdsp}, respectively). This can be understood in the limit $m_S^2 \rightarrow \infty$: Since the NMSSM potential differs from the MSSM potential then only in the quartic coupling, we expect differences in the fine-tuning expressions to also arise only in the dependence on this coupling. From the minimisation conditions for the potential, we find that $v_s \rightarrow 0$ for $m_S^2 \rightarrow \infty$. 
Setting $v_s=0$ in the NMSSM expressions for $m_h$ and $m_H$, we see that they differ from the corresponding MSSM expressions by $\lambda$-dependent terms. These terms arise from the $\lambda$-dependent contribution to the quartic coupling in the NMSSM potential.

Let us finally comment on the case where the decoupling of $h$ from $H$ and $s$ is achieved by making the mixing mass terms $m_{hH}^2$ and $m_{hs}^2$ small while keeping the diagonal matrix elements $m_H^2$ and $m_s^2$ at moderate values. We see from eq.~\eqref{mhHsqrelNMSSMdsp} that $m_{hH}^2$ vanishes if we choose $\lambda = 2\tilde{g}$ (see \cite{Delgado:2013zfa,Carena:2013ooa}). However, this gives $\lambda \sim 0.5$, meaning that $\lambda$ is not large enough to lift the Higgs mass to $126 \GeV$ (which requires $\lambda \sim 1$) without significant loop contributions from the stop sector (which in turn increase the fine-tuning). In addition, in absence of a UV completion which justifies the relation $\lambda \simeq 2\tilde{g}$, such a choice for $\lambda$ should be considered another type of tuning. Alternatively, we could consider large $\tan \beta$ since then $m_{hH}^2 \approx  v^2 (\lambda^2 - 4 \tilde{g}^2) / \tan \beta$ as follows from eq.~\eqref{mhHsqrelNMSSMdsp}. However, for $\lambda \sim 1$, electroweak precision tests restrict $\tan \beta \lesssim 4$ \cite{Barbieri:2006bg,Franceschini:2010qz,Gherghetta:2012gb}. Large $\tan \beta$ is thus not an option. This is an important difference compared to the MSSM. The dimensionful parameters $v_s$, $a_\lambda$, $\mu$ and $M$ that determine the mixing mass $m_{hs}^2$ according to eq.~\eqref{mshsqrelNMSSMdsp}, on the other hand, can not become arbitrarily small due to experimental and stability constraints. For example, the combination $\mu + \lambda v_s / \sqrt{2}$ that appears in eq.~\eqref{mshsqrelNMSSMdsp} is the effective $\mu$-term and therefore needs to be sufficiently large to satisfy collider constraints on charginos. In addition, the soft mass $a_\lambda$ must be relatively large to ensure the stability of the potential. The limit of vanishing $m_{hs}^2$ would therefore require an accidental cancellation among the various contributions in eq.~\eqref{mshsqrelNMSSMdsp}. Both of these types of tuning in the off-diagonal mass-mixing entries can be accounted for as tunings of the Higgs coupling ratios to SM-like values and can be studied with a measure analogous to the usual fine-tuning measure, 
$\smash{\Sigma_{u,d,V}\equiv \sqrt{\sum_\xi\left(\partial \log (r_{u,d,V}-1)/\partial \log  \xi\right)^2}}$, where $\xi$ are the input parameters.
This must be taken into account when assessing the naturalness of the model and would again increase the overall fine-tuning. It is not clear that this alternative way of obtaining SM-likeness once the overall fine-tuning is accounted for would provide any advantage from the viewpoint of naturalness compared to the decoupling via large masses for the heavy $CP$-even states. We thus do not consider this alternative way of obtaining SM-like couplings in our following analysis.

\subsection{The scale-invariant NMSSM}
\label{NMSSMscale-invariant}

\subsubsection{The fine-tuning measure}
\label{FT-NMSSMscale-invariant}

We shall now investigate the NMSSM with a scale-invariant superpotential
\be
W \, \supset \, \lambda S H_u H_d \, + \, \kappa S^3 \, .
\label{scaleinvNMSSM}
\ee
It has the advantage that it allows for the dynamical generation of the $\mu$-term and thereby solves the $\mu$-problem. Including soft terms and the $D$-term contribution, the potential for the electromagnetically neutral Higgs scalars $H^0_u$ and $H^0_d$ and the singlet scalar $S$ is given by
\begin{multline}
V =  (m_{H_{u}}^2+\lambda^2 |S|^2)|H^0_{u}|^2\, + \,(m_{H_{d}}^2+\lambda^2 |S|^2)|H^0_d|^2 \, + \,\lambda^2|H^0_{u}H^0_{d}|^2+m_S^2 |S|^2 \, + \, \kappa^2|S|^4\\
\, +\,  \left[\frac{a_{\kappa}}{3} S^{3} -(a_{\lambda} S+\lambda\kappa S^2)H^0_{u}H^0_{d} +\text{h.c.} \right]\, + \, \tilde{g}^2 \, (|H_u^0|^2 - |H_d^0|^2)^2 \, .
\label{fullpotential}
\end{multline}
Assuming $CP$-conservation, redefinitions can be used to make $\lambda$ and the vevs $\smash{v_s = \sqrt{2} \langle S \rangle}$ and $\langle H^0_{u,d} \rangle$ real and positive. All other parameters in the Higgs potential are also real but can have both signs. 
We shall again focus on the case $\lambda \sim 1$ so that the Higgs mass becomes $126 \GeV$ already at tree-level. We will neglect any loop corrections as discussed in sec.~\ref{NMSSMspmt}. 

The mass matrix elements for the $CP$-even states in the basis $(h,H,s)$ are reported in eq.~\eqref{MassMatrixElementsNMSSMsi} in appendix~\ref{CPeveneigensystem}. 
The Higgs potential \eqref{fullpotential} is determined by the parameters
$\{m_{H_u}^2,m_{H_d}^2, m_S^2, a_\lambda,a_\kappa, \lambda, \kappa, \tilde{g} \}\, .$
These parameters constitute the set $\xi$ over which we sum when calculating the fine-tuning measure \eqref{finetuningdef}. 
Using the expression for $d v/d \xi$ from appendix \ref{Finetuningexpressions} (see eq.~\eqref{dvdxiNMSSM}), we again obtain a formula for the fine-tuning measure which is partly given in terms of the Higgs-sector masses. In order to express the remaining pieces in terms of these masses, we first use the minimisation conditions $\ell_{h,H,s}=0$ for the potential (see~eq.~\eqref{minconditionsNMSSMsi} for the explicit expressions) to trade $\smash{m_{H_u}^2, m_{H_d}^2, m_S^2}$ for $\smash{v, \tan \beta, v_s}$. This allows us to express all quantities in terms of 
$\{v,\tan \beta, v_s, a_\lambda,a_\kappa, \lambda, \kappa, \tilde{g} \}\, $. In terms of these parameters, the elements of the Higgs mass matrix read
\begin{subequations}\label{massrelNMSSM}
\begin{align}
& m_h^2 \, = \, 2 \, \tilde{g}^2 v^2 \cos^2 \hspace*{-0.035cm} 2 \beta \, + \, \frac{1}{2} \lambda^2 v^2 \sin^2 \hspace*{-0.035cm} 2 \beta \label{mhsqrelNMSSM} \\
& m_H^2 \, = \,  \csc 2 \beta  \left( \sqrt{2} v_s a_\lambda + \kappa \lambda v_s^2 - \frac{v^2}{2} \sin^3 \hspace*{-0.035cm} 2 \beta 
   \left(\lambda^2- 4 \tilde{g}^2 \right)\right) \label{mHsqrelNMSSM}  \\
& m_s^2 \, = \, \frac{a_\kappa v_s}{\sqrt{2}} + 2 \kappa^2 v_s^2 +\frac{a_\lambda v^2 \sin 2 \beta }{\sqrt{8} \, v_s} \label{mssqrelNMSSM}  \\
& m_{h s}^2 \, = \, v \left(\lambda^2 v_s - \sin \beta  \cos \beta  (\sqrt{2} a_\lambda + 2 \kappa  \lambda  v_s) \right) \label{mshsqrelNMSSM} \\
& m_{h H}^2 \, = \, \frac{v^2}{4} \sin 4 \beta \, \left(4 \tilde{g}^2 - \lambda ^2\right) \label{mhHsqrelNMSSM}  \\
& m_{H s}^2 \, = \, \frac{v}{2} \, \cos 2 \beta \,  (\sqrt{2} a_\lambda + 2 \kappa  \lambda v_s) \label{msHsqrelNMSSM}  \, . 
\end{align}
\end{subequations}

By inverting eqs.~\eqref{mHsqrelNMSSM} to \eqref{mshsqrelNMSSM}, we can next trade $\smash{v_s, a_\lambda, a_\kappa}$ for the masses $\smash{m_H^2, m_s^2, m_{h s}^2}$. 
To this end, we solve eqs.~\eqref{mHsqrelNMSSM} and \eqref{mshsqrelNMSSM} for $a_\lambda$. Equating both results, we obtain an equation for $v_s$ which schematically reads $\smash{(2 \lambda - \kappa \sin 2 \beta) \, v_s^2  +\text{const.} \, v_s + \text{const.}=0}$, where we have explicitly given the prefactor of the $v_s^2$-term. 
Solving this for $v_s$, we find
\begin{subequations}\label{relalambdavs}
\begin{align}
&v_s \, = \, \frac{\pm  \sqrt{\mathcal{C}}  +  \, 2 m_{h s}^2  \csc 2 \beta}{2 v \lambda  \csc 2 \beta \, ( 2 \lambda - \kappa \sin 2 \beta)}  
\label{relvs} \\
&\begin{multlined}
a_\lambda \, = \, \frac{ \pm \sqrt{\mathcal{C}} \, (\lambda - \kappa \sin 2 \beta ) - 2  \lambda \, m_{hs}^2 \csc 2 \beta}{  \sqrt{2} v \, (2 \lambda - \kappa \sin 2 \beta)} \label{relalambda} \, ,
\end{multlined}
\end{align}
\end{subequations}
where 
\be
\mathcal{C} \, \equiv \, 4 \, m_{hs}^4 \csc^2 \hspace*{-0.035cm} 2 \beta + 2 \, \lambda  v^2   (2 \lambda - \kappa \sin 2 \beta ) \left(2 m_H^2-  v^2 \sin^2 \hspace*{-0.035cm} 2 \beta  \left(4 \tilde{g}^2-\lambda^2 \right)\right) \, . \label{cdef}
\ee
The two signs correspond to the two solutions from the quadratic equation for $v_s$. In the limit $2 \lambda \rightarrow \kappa \sin 2 \beta$, the prefactor of the $v_s^2$-term in that equation vanishes. 
Correspondingly, only one solution remains at this point (which appears as a pole in the above relations) for which we find
\begin{subequations}\label{limitrel}
\begin{align}
& v_s \, = \, \frac{v \sin^2 \hspace*{-0.035cm} 2 \beta \, \left(v^2 \sin^2 \hspace*{-0.035cm} 2 \beta \,  \left(4 \tilde{g}^2- \lambda^2 \right)- 2 m_H^2 \right)}{4 m_{hs}^2} \label{relvslimit}\\
& a_\lambda \, = \, \frac{\lambda^2 v^2 \sin 2 \beta \, \left(2 m_H^2 - v^2 \sin^2 \hspace*{-0.035cm} 2 \beta \, \left(4 \tilde{g}^2 - \lambda^2 \right)\right) - 2 m_{hs}^4 \csc \beta \, \sec \beta }{2 \sqrt{2} m_{hs}^2 v} \, .
\end{align}
\end{subequations}
The expression for $a_\kappa$ can be obtained from eq.~\eqref{mssqrelNMSSM} using the above relations for $v_s$ and $a_\lambda$. Note that $v_s$ and $a_\lambda$ do not depend on $m_s^2$, whereas $a_\kappa$ does. This reflects the fact that $a_\kappa$ only enters in the expression for $m_s^2$. 

Using eq.~\eqref{relalambdavs} and the relation for $a_\kappa$, we can express all quantities in terms of the parameters $\{v, \tan \beta, \lambda, \kappa,\tilde{g},  m_H, m_s, m_{h s}^2\}$.
In particular we find
\begin{equation} 
m_{Hs}^2 \, = \,   \frac{\left(\pm  \lambda \sqrt{\mathcal{C}} / 2 -  m_{hs}^2 \, ( \lambda \csc 2 \beta - \kappa ) \right) \cos 2 \beta }{2 \lambda - \kappa \sin 2 \beta} \, . \label{mHsrel}
\end{equation}

The two signs in the above relations are necessary in order to cover the entire parameter space when using the new basis $\smash{\{v, \tan \beta, \lambda, \kappa, \tilde{g}, m_H, m_s, m_{h s}^2\}}$. There are thus two possible values for $v_s$ and $a_\lambda$ according to eq.~\eqref{relalambdavs} (and the same applies to $a_\kappa$) for each combination of parameters in the new basis. On the other hand, there are restrictions on these parameters. To see this, recall that we can choose $v_s$ (together with $\lambda$ and $\langle H^0_{u,d} \rangle$) real and positive, whereas the other parameters in the Higgs potential can just be chosen real. This means that it is sufficient to only consider those combinations of parameters in the new basis that give a positive $v_s$ in \eqref{relvs}. In addition the fact that $v_s$ and $a_\lambda$ are real means that $\mathcal{C}$ has to be positive. This puts additional restrictions on these parameters. 
For example, for $\smash{2\lambda <  \kappa \sin 2 \beta}$, this implies that $m_H^2$ can not become arbitrarily large compared to $|m_{hs}^2|$.\footnote{In order to see how this arises in the old basis, let us for simplicity consider the case $m_{hs}^2=0$. The condition $\mathcal{C}>0$ then simplifies to
$$  m_H^2 \gtrless v^2 \sin^2 \hspace*{-0.035cm} 2 \beta \, (4 \tilde{g} - \lambda^2)/2 \, ,$$
where the two cases are for $2\lambda \gtrless \kappa \sin 2 \beta $. Let us reproduce this condition in the old basis. To this end, we 
solve $m_{hs}^2=0$ in eq.~\eqref{mshsqrelNMSSM} for $a_\lambda$ and use the result in eq.~\eqref{mHsqrelNMSSM} for $m_H^2$. This gives the following relation for $m_H^2$ in the old basis: $$ m_H^2 = \lambda v_s^2 \csc^2 \hspace*{-0.035cm} 2 \beta \, (2 \lambda - \kappa  \sin 2 \beta)  +   v^2 \sin^2 \hspace*{-0.035cm} 2 \beta \,  (4 \tilde{g}^2-\lambda^2 ) / 2 \, .$$ Depending on the sign of $2 \lambda - \kappa  \sin 2 \beta$, the first term is positive or negative. This relation thus reproduces the condition on $m_H^2$ that follows from the positivity of $\mathcal{C}$.} That $\mathcal{C}$ is positive is straightforward to see in the old basis: Using eqs.~\eqref{mHsqrelNMSSM} and \eqref{mshsqrelNMSSM} in the definition for $\mathcal{C}$, we find that 
\be 
\mathcal{C}= 2 v^2 \left(a_\lambda + \sqrt{2}\lambda^2v_s \csc 2 \beta \right)^2 \, .
\ee

Using eq.~\eqref{relalambdavs} and the relation for $a_\kappa$, we obtain an expression for the fine-tuning measure as a function of the new parameters. It is again too lengthy to be reported in this paper. We are interested in the limit of SM Higgs couplings, $\smash{m_H^2, m_s^2 \gg |m_{h H}^2|, |m_{h s}^2|}$. Similar to what we have discussed at the end of sec.~\ref{NMSSMspmt}, if we want to approach this limit by decreasing $m_{h H}^2$ and $m_{h s}^2$ (while keeping $m_H^2$ and $m_s^2$ at moderate values), we have to tune the parameters that determine these mixing masses. We will not consider this possibility further. Instead, we will focus on the case of large diagonal masses $m_H^2, m_s^2 \gg v^2, |m_{h s}^2|$. Expanding in $m_H^2$ and $m_s^2$, the expression simplifies considerably. We can distinguish two cases. For $m_s^2 \, v \gg m_H |m_{hs}^2|$, the fine-tuning measure is dominated by the term
\begin{multline}
\Sigma \, \approx \, f(\lambda, \kappa, \tan \beta,\tilde{g})  \, \,  \frac{m_H^2}{v^2} \, ,\\ \text{where} \, \, 
f(\lambda,\kappa, \tan \beta,\tilde{g}) \, \equiv \, \sqrt{ \frac{\cos 4 \beta  \left(\lambda^2- 13 \kappa^2 \right)- 52 \kappa  \lambda \sin 2 \beta + 13 \kappa^2 + 35 \lambda^2}{(\lambda^2 + 4 \tilde{g}^2\cot^2 \hspace{-0.035cm} 2\beta )^2 \, (2 \lambda - \kappa  \sin 2 \beta  )^2}} \, . \label{ftapprox1}
\end{multline}
For the opposite case $m_H |m_{hs}^2| \gg m_s^2 \, v$, the dominant term instead is
\begin{multline}
\Sigma \, \approx \, g(\lambda, \kappa, \tan \beta,\tilde{g}) \, \, \frac{ m_H^3}{ v^3 }\frac{|m_{hs}^2| }{ m_s^2}\, ,\\ \text{where} \, \, 
g(\lambda,\kappa,\tan \beta,\tilde{g}) \, \equiv \, 12 \,\kappa^2 \sqrt{\frac{ \sin^2 \hspace{-0.035cm} 2\beta   }{ \lambda^3(\lambda^2 +4 \tilde{g}^2 \cot^2 \hspace{-0.035cm} 2\beta)^2 \, (2  \lambda -\kappa  \sin 2 \beta )^3}} \, . \label{ftapprox2}
\end{multline}
We see that the fine-tuning measure does not increase but is suppressed for $m_s\gg m_H$. This is in analogy to what we have found in the last section. Again it can be understood from the fact that, in the limit $m_S^2,m_s^2 \rightarrow \infty$, the Higgs potential is just the MSSM potential plus an additional contribution to the quartic coupling. 

Note that eqs.~\eqref{ftapprox1} and \eqref{ftapprox2} are suppressed like $1/\tan^2\beta$ for $\tan\beta\gg 1$. Furthermore, we have checked that any term in the full expression for $\Sigma$ which is not suppressed by powers of $\tan\beta$ is suppressed by inverse powers of $m^2_H$. This shows that, similar to the MSSM, the increase in fine-tuning in the limit of SM Higgs couplings can be compensated by large $\tan\beta$. We emphasise, however, that $\tan \beta$ is restricted to the range $\tan \beta \lesssim 4$ due to electroweak precision tests for $\lambda \sim 1$  (so that the correct Higgs mass is obtained at tree-level).
But this observation can be relevant for implementations of the NMSSM with smaller values of $\lambda$ in which large values of $\tan\beta$ are less constrained. Note, however, that the fine-tuning would in turn increase due to stop corrections that are then necessary to raise the Higgs mass.

Notice that eqs.~\eqref{ftapprox1} and \eqref{ftapprox2} seemingly diverge for $2  \lambda \rightarrow \kappa  \sin 2 \beta$. This is an artefact of the expansion which breaks down near the pole. Using the relations in eq.~\eqref{limitrel} and the corresponding relation for $a_\kappa$, we find that the fine-tuning measure stays finite in this limit. Interestingly, however, its behaviour changes: Near the pole, the leading term is of order $m_H^4$.\footnote{This can can be understood as follows: First, recall that $\smash{\lambda \langle S \rangle =\lambda v_s / \sqrt{2}}$ is the effective $\mu$-term in the scale-invariant NMSSM. Next, notice from eqs.~\eqref{relvs} and \eqref{relvslimit} that in the limit of large $m_H$, one has $v_s\propto m_H$ away from the pole whereas $v_s \propto m_H^2$ at the pole. Let us assume that the NMSSM has the same $\mu$-dependence of the fine-tuning measure as the MSSM, eq.~\eqref{fine-tuning-large-tanbeta}. This then reproduces the $m_H$-dependence of the fine-tuning measure (and its change) both away from and at the pole.} Finally, recall that only when the combination $2  \lambda - \kappa  \sin 2 \beta $ is positive can $m_H^2$ become arbitrarily large compared to $|m_{hs}^2|$ (as follows from the positivity of $\mathcal{C}$). This ensures that the radicant in eq.~\eqref{ftapprox2} is always positive in the domain where the expansion is valid.

\subsubsection{A numerical scan of the parameter space}
\label{scansection}

It is useful to have an idea of the typical ranges for the parameters which determine the fine-tuning measure. 
We see from eqs.~\eqref{massrelNMSSM} that, in the limit of either large $v_s$, $a_\lambda$ or $a_\kappa$, the mass matrix elements go like
\begin{align}
m_H^2 \, &\propto \, v_s^2  \text{ or }  a_\lambda &
m_{hs}^2 \, &\propto \, v_s  \text{ or } a_\lambda \\\nonumber
m_s^2 \, &\propto \, v_s^2\, , \, a_\lambda \text{ or }  a_\kappa &
m_{Hs}^2 \, &\propto \, v_s  \text{ or }  a_\lambda \, ,
\end{align}
whereas $m_{hH}^2$ does not depend on $v_s$, $a_\lambda$ and $a_\kappa$ and is always of order the electroweak scale. 
The requirement that the theory has no tachyonic states limits the size of $a_\kappa$. Large $a_\lambda$, on the other hand, does not lead to decoupling of the singlet as $m_s^2$ and $m_{hs}^2$ grow similarly with it. For that reason, we expect that decoupling both $H$ and $s$ requires large $v_s$. 
Keeping only the dependence on dimensionful parameters, we find in the limit of large $v_s$ that
\be
m_H |m_{hs}^2| \, \sim \, v \, v_s^2 \, \sim \, v \, m_s^2 \, .
\ee
We should emphasise that, in practice, also $a_\lambda$ and $a_\kappa$ contribute to these quantities and make them either larger or smaller than the above estimate. A priori, either of the two contributions \eqref{ftapprox1} and \eqref{ftapprox2} to the fine-tuning measure can therefore dominate. 

\begin{figure}\centering
\begin{subfigure}{7cm}
\includegraphics[width=7cm]{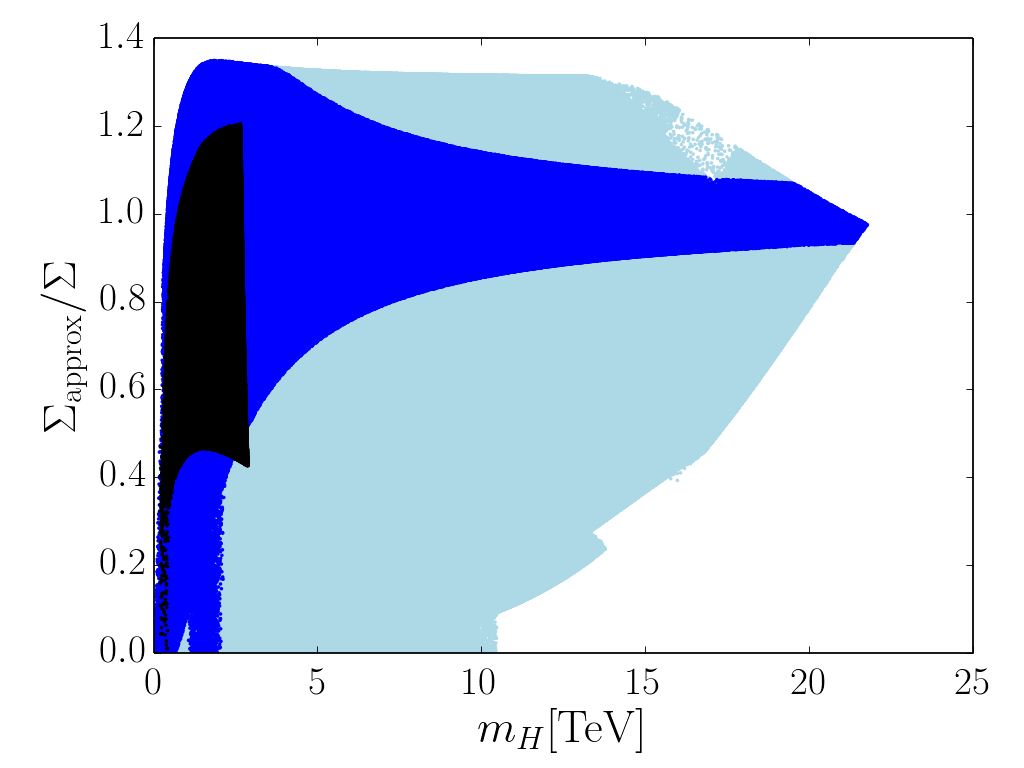}
\caption{}
\label{fig:FTratiovsmH}
\end{subfigure}
\hspace{1cm}
\begin{subfigure}{7cm}
\includegraphics[width=7cm]{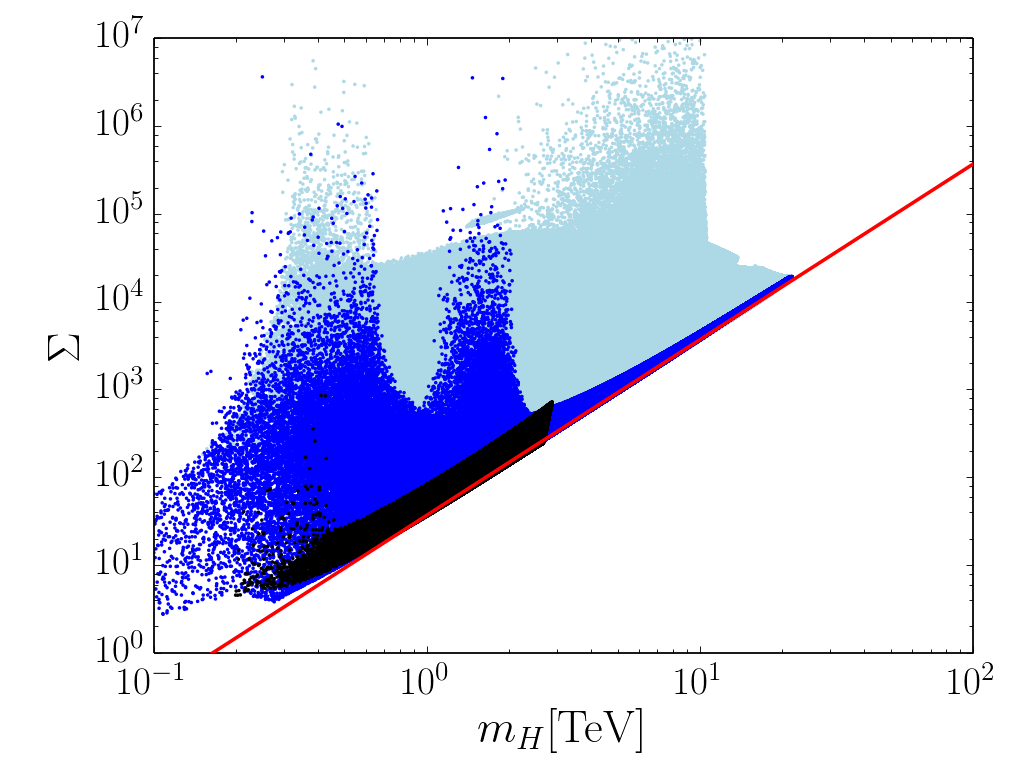}
\caption{}
\label{fig:FTvsmH}
\end{subfigure}
\caption{(a) The ratio of the analytic approximation given by \eqref{ftapprox1} and \eqref{ftapprox2} plus the third term mentioned in the text over the fine-tuning measure without approximation and (b) the fine-tuning measure without approximation, plotted against $m_H$. In both plots, the only constraint imposed on the light blue points is the absence of tachyonic states in the Higgs sector.  Dark blue points in addition satisfy $\smash{|m_{hs}^2| < (300 \GeV)^2}$. Black points have no tachyonic states and satisfy all experimental constraints implemented in \texttt{NMSSMTools 4.2.1} with the exception of $(g-2)_\mu$ and constraints related to dark matter.}
\end{figure}

In order to explore the viable parameter space and to verify our analytical findings, we have performed a numerical scan using the program \texttt{NMHDECAY}~\cite{Ellwanger:2004xm,Ellwanger:2005dv,Belanger:2005kh} contained in the package \texttt{NMSSMTools 4.2.1}. This also allows us to include the relevant experimental constraints and the dominant loop corrections from the effective potential. Concerning the latter, we are mainly interested in loop corrections from the Higgs sector to itself as these are connected to the Higgs couplings. In order to minimise loop corrections from other particles, we have therefore fixed the soft masses of gluinos and third-generation squarks to relatively small values which are still consistent with experiment ($m_{\tilde Q_3}=m_{\tilde u_3}=m_{\tilde d_3}=750 \GeV$, $A_t = 0$, $M_3=1.5 \TeV$). First- and second-generation squarks, sleptons and electroweak gauginos, on the other hand, give smaller loop corrections due to their small couplings. We have fixed their soft masses to the relatively large value of  $5 \TeV$. In the Higgs sector, we have fixed the dimensionless parameters to $\lambda=0.85$, $\kappa=0.8$ and $\tan\beta=3$ (we have also performed scans for different values of $\lambda$, $\kappa$ and $\tan \beta$ and found similar results).  Note that eq.~\eqref{ftapprox1} only depends on $m_H$ for fixed dimensionless parameters. We have then randomly varied the dimensionful parameters $v_s$, $a_\kappa$ and $a_\lambda$ within the following ranges: $100\GeV< \lambda v_s /\sqrt{2} < 8 \TeV$, $|a_\kappa / \kappa|<1 \TeV$, and $|a_\lambda / \lambda |<10 \TeV$. 

In order to verify our analytical approximation for the fine-tuning measure, we want to compare this approximation with the fine-tuning measure before expanding in $m_H$ and $m_s$ for the points found in the scan.
In this context, we should point out that \eqref{ftapprox1} and \eqref{ftapprox2} correspond to two different terms in an expansion of $\Sigma^2$. A third term in that expansion becomes important in the region where \eqref{ftapprox1} and \eqref{ftapprox2} are of comparable size. In the following, we take the square-root of these three terms arising in the expansion of $\Sigma^2$ as the approximation for $\Sigma$. In fig.~\ref{fig:FTratiovsmH}, we show the ratio of this approximation over the fine-tuning measure before expanding in $m_H$ and $m_s$ plotted against $m_H$. We find in the scan that generically $m_s \approx m_H$ within a factor of 3. Accordingly, we expect that the approximation becomes better for larger $m_H$. The scatter plot confirms this: Light blue points have no constraints imposed except for the absence of tachyonic states in the Higgs sector. For these points, the approximation agrees with the exact fine-tuning measure up to an $\mathcal{O}(1)$-factor for the upper range of $m_H$ found in the scan. Dark blue points in addition satisfy $\smash{|m_{hs}^2| < (300 \GeV)^2}$. For these points, the approximation works even better, the ratio being close to 1 for a large range of $m_H$ found in the scan. This improvement is related to the fact that the expansion leading to the approximation requires small $\smash{|m_{hs}^2|}$. In particular, note that our expansion in $m_H$ of $\mathcal{C}$ (which enters the fine-tuning measure via the relations for $v_s$, $a_\lambda$ and $a_\kappa$; see \eqref{cdef} for the explicit expression) requires that $\smash{|m_{hs}^2| \ll v \, m_H}$. Black points have no tachyonic states and fulfil all experimental constraints implemented in \texttt{NMSSMTools 4.2.1} with the exception of $(g-2)_\mu$ and constraints related to dark matter. In particular, the Higgs mass is within the range measured by ATLAS and CMS. We see that, for fixed $m_H$, the approximation works generically better for points satisfying the experimental constraints than for those that do not (though the range of $m_H$ found in the scan is smaller than for points not satisfying the experimental constraints). 

We show the fine-tuning measure (without an expansion in $m_H$ and $m_s$) plotted against $m_H$ in fig.~\ref{fig:FTvsmH}. The color code is the same as in fig.~\ref{fig:FTratiovsmH}. The straight line corresponds to our approximation \eqref{ftapprox1}. We see that, for points satisfying the experimental constraints, the fine-tuning measure grows like $m_H^2$. This shows that for these points, \eqref{ftapprox1} dominates the fine-tuning measure. We will use this fact in the next section.

\subsubsection{A lower bound on the fine-tuning measure}
We shall now connect the fine-tuning measure to the Higgs couplings in the scale-invariant NMSSM. 
Similar to our discussion for the MSSM, we approximately diagonalize the mass matrix for large $m_{H}$ and $m_s$ and find for the coupling ratios in the limit of small mixing (see appendix \ref{app:HiggsCouplings}):
\begin{subequations} \label{rurdrvNMSSM}
\begin{align}
r_u \, &\simeq \, 1 - \frac{m_{hH}^4}{2 m_H^4}  - \frac{m_{hs}^4}{2 m_s^4} + \cot \beta \, \left(\frac{m_{hH}^2}{m_H^2} -\, \frac{m_{hs}^2 \, m_{Hs}^2 }{m_H^2 m_s^2} +\frac{m_{h}^2 m_{hH}^2}{m_H^4}\right)
 \label{ruNMSSM} \\
r_d \, &\simeq  \, 1 - \frac{m_{hH}^4}{2 m_H^4}  - \frac{m_{hs}^4}{2 m_s^4}  - \tan \beta \, \left(\frac{m_{hH}^2}{m_H^2} -\, \frac{m_{hs}^2 \, m_{Hs}^2 }{m_H^2 m_s^2} +\frac{m_{h}^2 m_{hH}^2}{m_H^4}\right)
  \label{rdNMSSM}  \\
r_V \, &\simeq  \, 1 - \frac{m_{hH}^4}{2 m_H^4}  - \frac{m_{hs}^4}{2 m_s^4} 
 \label{rvNMSSM} \, .
\end{align}
\end{subequations}

\begin{figure}[t]
\begin{subfigure}{.47\linewidth}
\includegraphics[width=\linewidth]{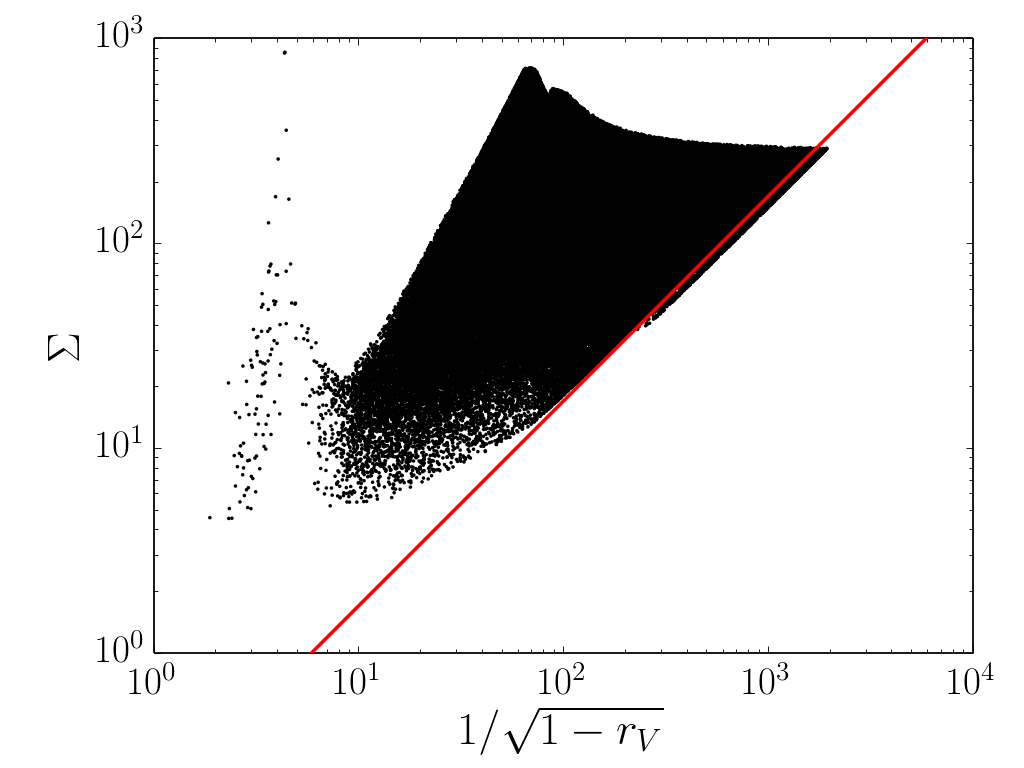}
\caption{ \label{FT-rv}}
\end{subfigure}
\hspace*{.3cm}
\begin{subfigure}{.47\linewidth} 
\includegraphics[width=\linewidth]{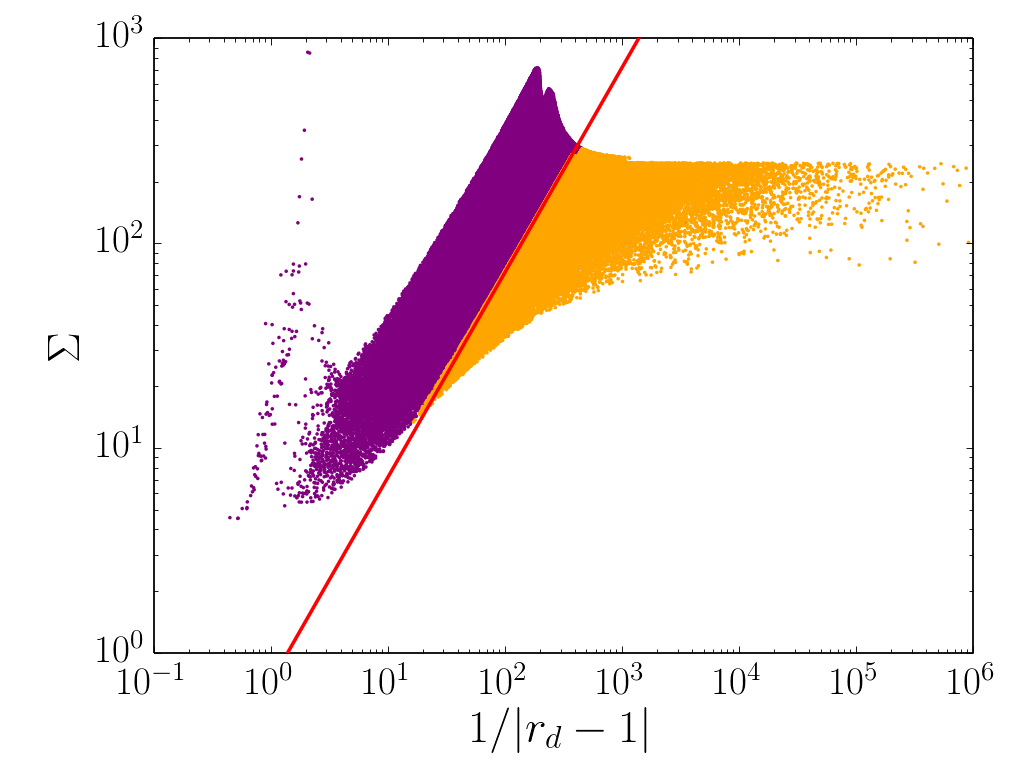}
\caption{\label{FT-rd}}
\end{subfigure}
\caption{The fine-tuning measure plotted against (a) $\smash{1/\sqrt{1-r_V}}$ and (b) $\smash{1/|r_d-1|}$ for the points from our scan that have no tachyonic states and satisfy the experimental constraints. In (b), purple points have $\Omega_d >0$, whereas this quantity is negative for orange points. The straight lines are the corresponding bounds from \eqref{finetuningvscouplingsrv} and \eqref{finetuningvscouplings}. If we plot the combined fine-tuning $\Sigma \cdot \Sigma_d$ instead of $\Sigma$, all orange points pile up at the red line. This shows that \eqref{finetuningvscouplings} is a lower bound on the combined fine-tuning.}
\label{FT-r}
 \end{figure}

Let us assume that the LHC and possibly the ILC do not observe any deviations in the Higgs couplings from the SM.
Given a certain precision in the measurement of the coupling ratios $r_{u,d,V}$, we can use our results to estimate the fine-tuning which is necessary in order to achieve this level of SM-likeness in the scale-invariant NMSSM. 
Recall from eqs.~\eqref{ftapprox1} and \eqref{ftapprox2} that the fine-tuning measure grows with $m_H$ but not with $m_s$. 
In order to estimate the fine-tuning for given limits on the coupling ratios, we thus need an expression for $m_H$ in terms of the latter.  
Note that the terms of order $m_H^{-4}$ in the relations for $r_{u,d}$ are negligible since $\tan \beta$ is not large. Using the resulting simpler relations for $r_{u,d,V}$ and solving for $m_H$, we find
\begin{subequations} \label{mHrels}
\begin{align}
m_H^2 \, & \simeq \, \frac{m_{hH}^2\,\cot \beta}{|r_u - 1|} \, \left(1 + \Omega_u \right) \label{mHestimateRU} \\
m_H^2 \, & \simeq \, \frac{m_{hH}^2\,\tan \beta}{|r_d - 1|} \, \left(1 + \Omega_d  \right)\label{mHestimateRD} \\
m_H^2 \, & \simeq \, \frac{|m_{hH}^2|}{\sqrt{2 (1-r_V) -\frac{m_{hs}^4}{m_s^4} }}\label{mHestimateRV} \, ,
\end{align}
\end{subequations}
where
\begin{subequations}
\begin{align}
\Omega_{u} \, \equiv \,  \frac{|r_{u} -1|}{r_{u} - 1 +\frac{m_{hs}^4}{2 m_s^4}} \left( 1 - \frac{m_{hs}^2 m_{Hs}^2 }{m_s^2 m_{hH}^2} \right) - 1\\
\Omega_{d} \, \equiv \,  \frac{|r_{d} -1|}{ 1- r_{d} - \frac{m_{hs}^4}{2 m_s^4}} \left( 1 - \frac{m_{hs}^2 m_{Hs}^2 }{m_s^2 m_{hH}^2} \right) - 1 \, .
\end{align}
\end{subequations}
These relations can be used to determine the value of $m_H$ which is necessary in order to satisfy given limits on $r_{u,d,V}$.
Note, however, their dependence in particular on $m_s$ and $m_{hs}^2$ which results from the singlet admixture to the Higgs. These masses are input parameters in the basis $\smash{\{v, \tan \beta, \lambda, \kappa,  \tilde{g},  m_H, m_s, m_{h s}^2\}}$ and undetermined ($m_{hH}^2$ and $m_{Hs}^2$, on the other hand, are determined via eqs.~\eqref{mhHsqrelNMSSM} and \eqref{mHsrel}).
For the $r_V$-dependent expression \eqref{mHestimateRV}, the admixture always increases $m_H$ and thus the fine-tuning. In the limit of vanishing mixing, $\smash{m_{hs}^2/m_s^2}\rightarrow 0$, we therefore obtain the smallest viable $m_H$ for a given bound on $r_V$. Since \eqref{ftapprox2} also vanishes in this limit, \eqref{ftapprox1} yields a lower bound on the fine-tuning measure,
\be
\Sigma \, \gtrsim \,  \frac{1}{4} \, (4\tilde{g}^2 -\lambda^2 ) \, f(\lambda,\kappa,\tan \beta,\tilde{g}) \, 
\frac{\sin 4 \beta}{ \sqrt{2 (1-r_V)}} \, ,\label{finetuningvscouplingsrv}
\ee
where the function $f$ is defined in \eqref{ftapprox1}. We see that the fine-tuning diverges in the SM limit $r_{V} \rightarrow 1$ as expected. This can not be compensated with large $\tan \beta$ since electroweak precision tests limit $\tan \beta \lesssim 4$ for $\lambda \sim 1$ (so that the correct Higgs mass is obtained at tree-level). This is an important difference compared to the MSSM. Even though smaller $\lambda$ allows for larger $\tan \beta$, the then necessary stop corrections to raise the Higgs mass would in turn increase the fine-tuning. 
Note that the undetermined quantities $\lambda, \kappa, \tan \beta$ in \eqref{finetuningvscouplingsrv} are numbers of order 1. Even if we do not know  their values, \eqref{finetuningvscouplingsrv} therefore still gives an order-of-magnitude estimate for the fine-tuning.
As a caveat, we point out that \eqref{finetuningvscouplingsrv} is applicable only for sufficiently small $r_{V}$ (since only then the corresponding $m_H$ is large enough to justify the expansion in $m_H$). We show a scatter plot of $\Sigma$ versus $\smash{1/\sqrt{1-r_V}}$ for the points from our scan in fig.~\ref{FT-rv}. The straight line corresponds to \eqref{finetuningvscouplingsrv}, confirming that it is a lower bound on the fine-tuning measure.\footnote{Note that there are points which are slightly below the straight line. This is due to the fact that \eqref{finetuningvscouplingsrv} was derived using only the leading contribution \eqref{ftapprox1} to the fine-tuning measure. However, the fine-tuning for the points was calculated using the exact expression for the fine-tuning measure which differs from \eqref{ftapprox1} by small corrections.}

The admixture with the singlet affects the required value of $m_H$ from \eqref{mHestimateRU} and \eqref{mHestimateRD} via the quantities $\Omega_{u,d}$.
It again increases $m_H$ and thereby the fine-tuning if these quantities are positive. To see this, note that $m_{hH}^2$ is positive for the parameter range of interest to us and that $m_H^2$ needs to be positive. Among points in parameter space with positive $\Omega_{u}$ or $\Omega_d$, those for which this quantity vanishes therefore require the smallest $m_H$ and have the smallest fine-tuning. Note that $\Omega_{u,d}\rightarrow 0$ in the limit of vanishing mixing, $m_{hs}^2/m_s^2 \rightarrow 0$. Since \eqref{ftapprox2} also vanishes in this limit, \eqref{ftapprox1} yields a \emph{lower} bound on the fine-tuning measure for these points:
\be
\Sigma \, \gtrsim \,  \frac{1}{4} \, (4\tilde{g}^2 - \lambda^2 ) \, f(\lambda,\kappa,\tan \beta,\tilde{g}) \, \cdot
\begin{cases}
\frac{\cot \beta \, \sin 4 \beta}{|r_u-1|} \\
\frac{\tan \beta \, \sin 4 \beta}{|r_d-1|} \, .
\label{finetuningvscouplings}
\end{cases}
\ee
The admixture lowers the required value of $m_H$ from \eqref{mHestimateRU} and \eqref{mHestimateRD}, on the other hand, for negative $\Omega_{u,d}$. Accordingly, \eqref{finetuningvscouplings} becomes an \emph{upper} bound on the fine-tuning for points in parameter space for which these quantities are negative. This is confirmed in fig.~\ref{FT-rv}, where we show a scatter plot of $\Sigma$ versus $\smash{1/|r_d-1|}$. Orange points have negative $\Omega_d$, whereas for purple points this quantity is positive. The straight line corresponds to the $r_d$-dependent bound from \eqref{finetuningvscouplings}, separating orange and purple points as expected. 

For $\Omega_{u,d}$ close to $-1$ (the smallest value consistent with the positivity of $m_H^2$), the required value of $m_H^2$ from \eqref{mHestimateRU} and \eqref{mHestimateRD} can become arbitrarily small. This is due to an accidental cancellation in the relations for the coupling ratios which allows the $|r_{u,d}-1|$-limits to be already satisfied with small $m_H$. However, the tuning corresponding to this cancellation should be taken into account when assessing the naturalness of the model. Let us estimate the amount of tuning. Without any accidental cancellation, the required value of $m_{H}$ for a given limit on $r_u$ would instead be $\smash{m_{H, {\rm nat.}}^2 \sim \cot \beta \,  m_{hH}^2 / |r_u-1|}$ and similarly for  $r_d$. We can take $\smash{\Sigma_{u,d} \sim m_{H, {\rm nat.}}^2 /m_{H}^2}$ as an estimate for the corresponding amount of tuning for points with negative $\Omega_{u,d}$ (whereas for positive $\Omega_{u,d}$ we set $\Sigma_{u,d}=1$, since there is no corresponding tuning) and define the product $\smash{\Sigma \cdot \Sigma_{u,d}}$ to measure the combined fine-tuning. Now recall that, for the points in our scan which satisfy the experimental constraints, the fine-tuning measure is well described by \eqref{ftapprox1} with a quadratic dependence on $m_H$. Furthermore, note that \eqref{mHestimateRU} and \eqref{mHestimateRD} reduce to $m_{H, {\rm nat.}}^2$ in the limit $\Omega_{u,d} \rightarrow 0$ in which \eqref{finetuningvscouplings} was derived. If we replace the fine-tuning measure $\Sigma$ by a combined measure $\smash{\Sigma \cdot \Sigma_{u,d}}$, \eqref{finetuningvscouplings} therefore gives a lower bound also for negative $\Omega_{u,d}$. We will accordingly use \eqref{finetuningvscouplings} as a lower bound also for points in parameter space with negative $\Omega_{u,d}$.

Note that \eqref{finetuningvscouplings} is in fact quite conservative as a lower bound: In our scan, we find that $m_s \approx m_H$ to within a factor of 3. This means that the limit $m_{hs}^2/m_s^2 \rightarrow 0$ requires $m_{hs}^2 \rightarrow 0$. As discussed at the end of sec.~\ref{FT-NMSSMscale-invariant}, this in turn requires an accidental cancellation among the various soft masses that determine $m_{hs}^2$. Again this additional tuning should be taken into account. Since \eqref{finetuningvscouplings} is saturated for points with $m_{hs}^2/m_s^2 =0$, the resulting combined fine-tuning would satisfy an even more stringent lower bound. 

\begin{figure}[t]
\begin{subfigure}{.47\linewidth}
\includegraphics[width=\linewidth]{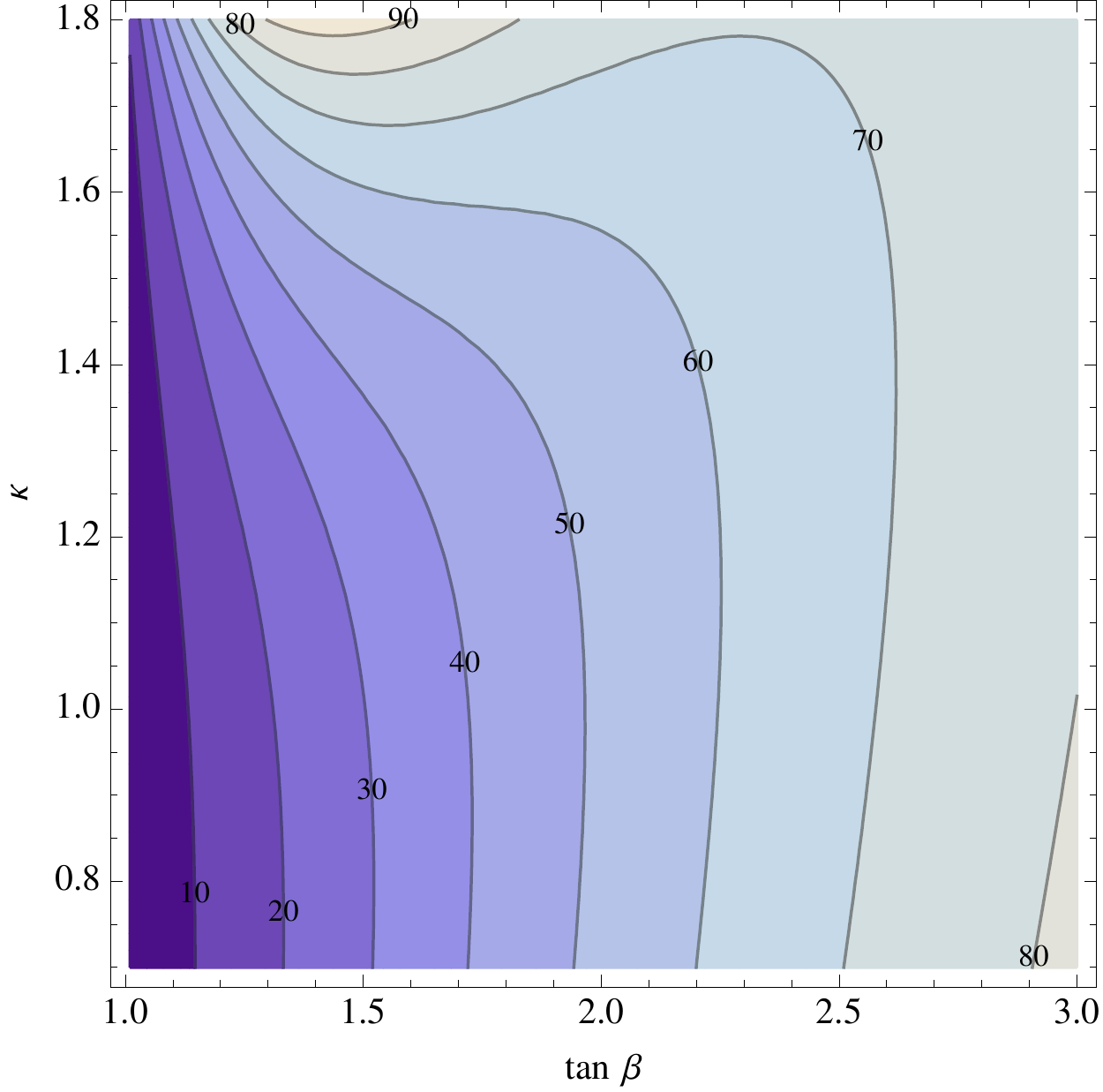}
\caption{ \label{fig:CouplingFineTuning1a}}
\end{subfigure}
\hspace*{.3cm}
\begin{subfigure}{.47\linewidth} 
\includegraphics[width=\linewidth]{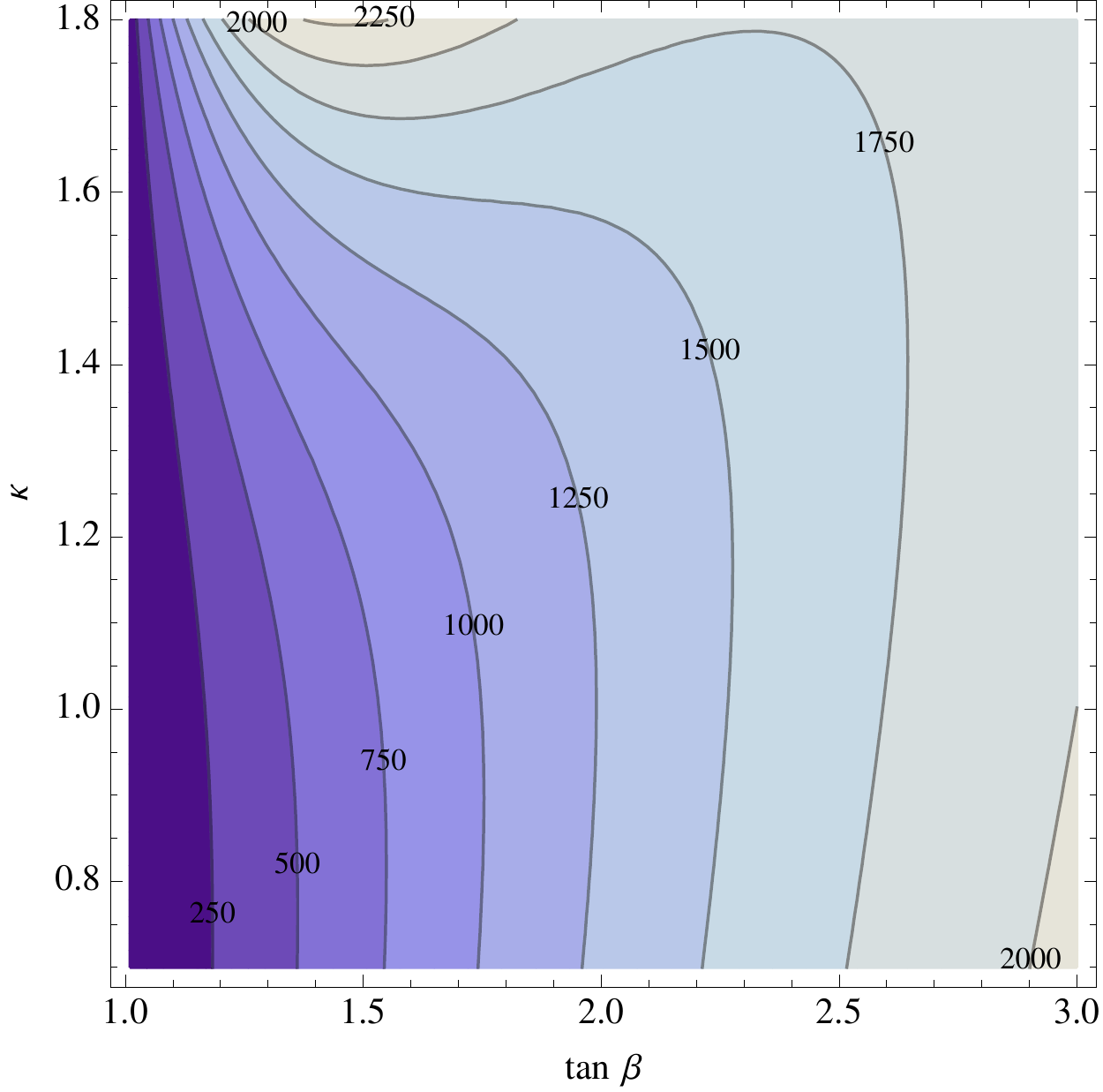}
\caption{\label{fig:CouplingFineTuning1b}}
\end{subfigure}
\caption{Contours of the fine-tuning measure $\Sigma$ in the $\kappa$-$\tan \beta$ plane for (a) the LHC at $14 \TeV$ with integrated luminosity of 300 fb$^{-1}$ and (b) the ILC at $1 \TeV$ with integrated luminosity of 2500 fb$^{-1}$.}
\label{fig:CouplingFineTuning1}
\end{figure}

\subsubsection{Implications of future Higgs coupling measurements for naturalness}
The precision with which the Higgs coupling can be measured at the LHC at $14 \TeV$ was estimated in~\cite{Peskin:2013xra}. We will focus on their `scenario~1' which makes the conservative assumption of no improvement over time in theoretical and systematic errors compared to current values. We use the $1 \sigma$-estimates from their fit for 300 fb$^{-1}$ integrated luminosity and allowing for invisible decay modes of the Higgs. Note that this fit allows for different coupling ratios to tops and charms, to bottoms and tauons and to $W$- and $Z$-bosons. In supersymmetry at tree-level, on the other hand, all up-type fermions have the same coupling ratio and similarly for down-type fermions and vectors (see appendix~\ref{app:HiggsCouplings}). Redoing the fit with these additional constraints would likely lead to somewhat higher estimates for the achievable precisions. Note that the said fit from \cite{Peskin:2013xra} imposes the condition $r_V<1$ (which is satisfied in our case, see eq.~\eqref{rvNMSSM}). 

Since the achievable precision at the LHC is not sufficient to justify an expansion in large $m_H$ (see the caveat below eq.~\eqref{finetuningvscouplingsrv}), we will use the full expression for the fine-tuning measure in the basis $\{v, \tan \beta, \lambda, \kappa,  \tilde{g},  m_H, m_s, m_{h s}^2 \}$ which we derive using Mathematica. 
In order to obtain a conservative estimate of the fine-tuning, we again focus on the limit of vanishing mixing with the singlet, $\smash{m_{hs}^2/m_s^2 \rightarrow 0}$. Since the fine-tuning measure depends on $m_s$ and $m_{hs}^2$ separately, we need to fix one of these masses in addition to the ratio $m_{hs}^2/m_s^2$. 
Since in our scan $m_s$ always lies in the range $\smash{m_H/3 \lesssim m_s \lesssim m_H}$ if all experimental constraints are satisfied, we choose $m_s= m_H / 2$ and accordingly set $m_{hs}^2=0$. 
Taking the corresponding tuning in $m_{hs}^2$ into account would increase our estimate for the fine-tuning. Nevertheless, the estimate that we present here remains conservative in the sense of being a lower bound.

We will assume that loop corrections to the Higgs couplings from superpartners  are small compared to the tree-level effect from the Higgs admixture.
We calculate $m_H$ from eq.~\eqref{mHrels} with $m_{hs}^2/m_s^2=0$ and take the smallest value which ensures that all three coupling ratios are within the ranges given in \cite{Peskin:2013xra}. 
Comparing the achievable precisions in the three coupling ratios, we find that the most stringent requirement on $m_H$ comes from down-type couplings. 
We fix $\lambda$ by the requirement that eq.~\eqref{HiggsMassNMSSM} gives the correct Higgs mass.
In fig.~\ref{fig:CouplingFineTuning1a}, we plot contours of the fine-tuning measure as a function of the remaining two free parameters $\kappa$ and $\tan \beta$.
The range of $\kappa$ is motivated by the range found in the scan in \cite{Gherghetta:2012gb}.
The range of $\tan \beta$, on the other hand, is limited by electroweak precision tests, since the neutralino contribution to the $T$-parameter increases with growing $\lambda$ and $\tan \beta$ away from the custodial symmetry limit, $\tan\beta = 1$. 
 We take the range of $\lambda$ and $\tan \beta$ found in the scan in \cite{Gherghetta:2012gb} to estimate the resulting limit on $\tan \beta$ vs.~$\lambda$. In addition, we have to ensure that $\lambda$ does not hit a Landau pole at too low scales. The coupling grows from $\lambda \approx 1$ at $\tan \beta =1$ to $\lambda \approx 1.7$ at $\tan \beta =3$. For these values, the Landau pole occurs well above $20 \TeV$. The growth of $\lambda$ with $\tan \beta$ is necessary in order to obtain the correct Higgs mass at tree-level (cf.~eq.~\eqref{mhsqrelNMSSM}). Even though the fine-tuning decreases with $\tan \beta$ (see the discussion below eq.~\eqref{ftapprox2}), the growth in $\lambda$ counteracts this so that overall the fine-tuning increases with $\tan \beta$ in the plots and eventually flattens out.

In \cite{Peskin:2013xra}, achievable precisions are also presented for the high-luminosity LHC with 3000 fb$^{-1}$. 
Using these potential limits on the coupling ratios, the increase in fine-tuning compared to 300 fb$^{-1}$ amounts to only about 50$\%$. The reason is that the precision for down-type couplings improves by only about $30\%$. In addition, \cite{Peskin:2013xra} gives the achievable precisions for the more optimistic `scenario~2', where theoretical errors are assumed to be halved and systematic errors are assumed to decrease as the square root of the integrated luminosity. Using the precisions for `scenario 2' with 3000 fb$^{-1}$, we find that the fine-tuning increases by more than a factor 3 compared to `scenario~1' with 300 fb$^{-1}$.

We see that, based on coupling measurements alone, the LHC can probe the naturalness of the NMSSM down to the few-percent level or better for a large range in the $\kappa$-$\tan \beta$ plane. This should be compared with the level of fine-tuning that is driven by collider constraints on stops and gluinos. In \cite{Gherghetta:2012gb}, we have found that under certain assumptions made to optimise the naturalness of the model -- a large coupling $\lambda$, a low messenger scale and a split sparticle spectrum (see \cite{Gherghetta:2012gb} for more details) -- stops and gluinos as heavy as, respectively, $1.2 \TeV$ and $3 \TeV$ can still be consistent with fine-tuning at the $5\%$-level.\footnote{Note, however, that the fine-tuning can be reduced up to a factor of 2-3 by a further doubling of the superpartner content \cite{Craig:2013fga}.} The exclusion reach for stops and gluinos at the $14 \TeV$ LHC with the ATLAS detector was estimated in \cite{ATL-PHYS-PUB-2013-002}. Even with 3000 fb$^{-1}$ integrated luminosity (though assuming a simplified model for stop and gluino decays), the exclusion reach remains below $1.1 \TeV$ for stops and $2.7 \TeV$ for gluinos. This means that searches for coloured sparticles and measurements of the Higgs couplings at the LHC may probe the naturalness of the NMSSM at a comparable level.\footnote{The model-building assumptions made in \cite{Gherghetta:2012gb}, however, may potentially require a relatively baroque model to be realised. Simpler models will likely require a larger amount of fine-tuning once they satisfy the collider constraints~\cite{Arvanitaki:2013yja}.}

This would change with the advent of the ILC. No improvements over the projected LHC limits on stops and gluinos will be possible with this collider. The precision in the Higgs coupling measurements, on the other hand, will be significantly improved over the LHC. 
We again use the precision estimates from \cite{Peskin:2013xra}. Note that their fit for the ILC imposes neither the constraint $r_V < 1$ nor that there are universal coupling ratios for up-type fermions, down-type fermions and vectors. Redoing the fit with these constraints would likely lead to higher precision estimates. We assume that the ILC uses information on the ratio of branching fractions BR$\smash{(h \rightarrow \gamma \gamma)}$/BR$\smash{(h \rightarrow Z Z^*)}$ from the LHC as discussed in \cite{Peskin:2013xra}. The most stringent constraint on $m_H$ again arises from down-type couplings. Different configurations of the ILC are considered in \cite{Peskin:2013xra}, the initial configuration being at $250 \GeV$ collision energy with 250 fb$^{-1}$ integrated luminosity.
The improvement in the precision for down-type couplings for this configuration compared to the LHC at $14 \TeV$ with 300 fb$^{-1}$ amounts to about 40 $\%$. Since the fine-tuning grows inversely linear with $|1-r_d|$ according to eq.~\eqref{finetuningvscouplings}, the required fine-tuning to satisfy the limits on the coupling ratios increases by only about 60 $\%$ compared to the LHC at $14 \TeV$ with 300 fb$^{-1}$. We therefore show a plot only for the final configuration at $1 \TeV$ with 2500 fb$^{-1}$. 
We again fix $\lambda$ via the Higgs mass and plot contours of the fine-tuning measure as a function of the remaining free parameters $\kappa$ and $\tan \beta$ in fig.~\ref{fig:CouplingFineTuning1b}. The coupling $\lambda$ grows from $\lambda \approx 1$ at $\tan \beta =1$ to $\lambda \approx 1.6$ at $\tan \beta = 3$. 
We see that at $1 \TeV$ and with 2500 fb$^{-1}$, the ILC could probe the fine-tuning down to the per-mille level. Precision measurements of the Higgs couplings can accordingly become an important tool to constrain the naturalness of the NMSSM.

\section{Conclusion}
\label{sec:conclusions}

The SM-like values of the Higgs couplings to fermions and gauge bosons measured at the LHC introduce a new source of fine-tuning in supersymmetric models. In the MSSM, the particle identified with the ${126 \GeV}$ Higgs boson is a mixture of a $CP$-even state with SM Higgs couplings and an additional,  
heavier $CP$-even state. Increasing the mass of this heavy $CP$-even state gives rise to more SM-like couplings, but also causes a further increase in the tuning of the electroweak vev. However this tuning can be offset by large $\tan\beta$ so that the overall tuning is not necessarily increased beyond that which arises from the usual contributions of stops, gluinos and Higgsinos (or any other contribution from the $D$-term sector). 

For the NMSSM, with an additional singlet field, this is no longer the case.  An order-one Higgs-singlet coupling is sufficient to obtain the $126 \GeV$ Higgs mass at tree-level. This large coupling, however, enhances the violation of custodial symmetry so that constraints from the $T$-parameter restrict $\tan\beta$ to small values. This means that the increase in fine-tuning from decoupling of the heavy $CP$-even states cannot be offset with large $\tan\beta$. Thus the NMSSM has a new source of tuning from Higgs coupling measurements. Even though the SM Higgs can now mix with two $CP$-even states the fine-tuning measure does not grow with the singlet mass, but at leading order grows quadratically with the mass of the $CP$-even state already present in the MSSM.

We derive a relation between the mass of this state and deviations in the Higgs couplings from SM values. In combination with our expression for the fine-tuning measure this relation can be used to see what future collider measurements will teach us about the naturalness of supersymmetric models. In particular, we consider the scale-invariant NMSSM with an order one Higgs-singlet coupling and derive a lower bound on the tuning.
 At Run-II of the LHC with 300 ${\rm fb}^{-1}$ the measurement of the Higgs 
couplings will probe the naturalness of this model at the few-percent level, roughly comparable with the tuning from current direct limits on the superpartner 
masses (assuming a model with a low messenger scale ($20 \TeV$) and split families). Instead at a $1 \TeV$ ILC with 2500 ${\rm fb}^{-1}$, more precise measurements of the Higgs couplings will probe naturalness at the per-mille level, corresponding to an approximately factor of 30 increase in the tuning. This is beyond any tuning derived from direct superpartner limits, so that the naturalness of supersymmetric models will be definitively tested at a future ILC.

\section*{Acknowledgements}
We thank Marco Farina and Sogee Spinner for helpful discussions. This work was supported in part by the Australian Research Council. 
TG was also supported in part by the DOE grant DE-FG02-94ER-40823. TG and BvH thank the Galileo Galilei Institute for Theoretical Physics in Florence for hospitality and the INFN for partial support during the completion of this work. BvH also thanks the Fine Theoretical Physics Institute at the University of Minnesota for hospitality and partial support. We acknowledge the use of \texttt{matplotlib}~\cite{Hunter:2007} and \texttt{ipython}~\cite{PER-GRA:2007}.

\appendix

\section{The $CP$-even Higgs mass matrix}
\label{CPeveneigensystem}

\subsection{MSSM}
In the MSSM and its $D$-term extension discussed in sec.~\ref{D-term-section}, the mass matrix for the $CP$-even states $h$ and $H$ is given by
\be
\mathcal{M}^2=\begin{pmatrix}
m_h^2 & m_{hH}^2 \\ m_{hH}^2 &  m_H^2
\end{pmatrix} \, ,
\ee
where
\begin{subequations}
\label{potentialbasis2b}
\begin{align}
& m_h^2 \,  = \, m_{H_d}^2\cos^2\hspace*{-0.035cm} \beta + m_{H_u}^2\sin^2 \hspace*{-0.035cm} \beta + 3 \tilde{g}^2 v^2\cos^2 \hspace*{-0.035cm} 2\beta + \mu^2  - B_\mu  \sin 2 \beta \\
& m_H^2 \, = \,m_{H_d}^2\sin^2\hspace*{-0.035cm} \beta+ m_{H_u}^2\cos^2 \hspace*{-0.035cm} \beta +  3 \tilde{g}^2 v^2 \sin^2 \hspace*{-0.035cm} 2\beta -\tilde{g}^2 v^2 +  \mu^2 + B_\mu \sin 2 \beta \\
& m_{hH}^2 \,  = \frac{1}{2} \left((m_{H_d}^2 - m_{H_u}^2) \sin 2 \beta +   3 \tilde{g}^2 v^2 \sin 4 \beta + 2 B_\mu \cos 2 \beta \right) \label{mHmathcalH}\, .
\end{align}
If loop corrections from the stop sector are important (in particular if they are used to raise the quartic coupling instead of additional $D$-terms), these matrix elements have sizeable corrections arising from the Coleman-Weinberg potential (cf.~eqs.~\eqref{stoploopmhsq} to \eqref{stoploopmhHsq}). 

\end{subequations}

\subsection{NMSSM}
In the NMSSM, the mass matrix for the $CP$-even states $h$, $H$ and $s$ is given by
\begin{equation}
\mathcal{M}^2\equiv\begin{pmatrix}
m_{h}^2 & m_{hH}^2 & m_{hs}^2\\
m_{hH}^2 & m_H^2 & m_{Hs}^2\\
m_{hs}^2 & m_{Hs}^2 & m_s^2 
\end{pmatrix}\, .
\end{equation}
The expressions for the matrix elements depend on the choice of the superpotential. 
For the superpotential with mass terms \eqref{spmt} (parameterised by $\mu, M, \lambda$), the matrix elements are given by
\begin{subequations}\label{MassMatrixElementsNMSSMdsp}
\begin{align}
&\begin{multlined}\label{MassMatrixElementsNMSSMdspmh}
m_h^2 \,  = \,\frac{1}{8} \left(  24 \tilde{g}^2 v^2\cos^2 \hspace*{-0.035cm} 2\beta +6 \lambda^2 v^2 \sin^2 \hspace*{-0.035cm} 2\beta+ 8\mu^2 + 8 m_{H_d}^2\cos^2 \hspace*{-0.035cm} \beta + 8m_{H_u}^2\sin^2\hspace*{-0.035cm} \beta    \right. \\ \left.  -  4 \sin 2 \beta \,  \left(\sqrt{2} v_s (a_\lambda + \lambda M) + 2 B_\mu \right)+    + 4 \lambda^2 v_s^2 + 8 \sqrt{2} \lambda  \mu  v_s \right) \end{multlined}\\
& \begin{multlined}  \label{MassMatrixElementsNMSSMdspmH}
 m_H^2 \, = \, \frac{1}{8} \left(4 \left( \sin 2 \beta  \left( \sqrt{2} a_\lambda  v_s + 2 B_\mu \right) + \tilde{g}^2 v^2 + 2 \mu^2 \right) + 3 v^2 \cos 4 \beta  \left( \lambda^2 - 4 \tilde{g}^2 \right) \right. \\ \left. + 8 \sqrt{2} \lambda  v_s (\mu + M \sin \beta \, \cos \beta ) + 8 m_{H_u}^2\cos^2 \hspace*{-0.035cm} \beta +8 m_{H_d}^2 \sin^2 \hspace*{-0.035cm} \beta + \lambda^2 \left(v^2 + 4 v_s^2 \right)\right)
 \end{multlined}\\
 & m_s^2 \, = \, m_S^2 + M^2 +\frac{1}{2} \lambda^2 v^2  \label{MassMatrixElementsNMSSMdspms} \\
 & m_{hs}^2 \, = \, v\,  (\sqrt{2} \lambda \mu +\lambda^2   v_s - \sqrt{2} \sin \beta  \,  \cos \beta  \, ( a_\lambda + \lambda  M))  \label{MassMatrixElementsNMSSMdspmhs} \\
 & m_{Hs}^2 \, = \, \frac{v}{\sqrt{2}}  \cos 2 \beta  (a_\lambda + \lambda  M) \, .\label{MassMatrixElementsNMSSMdspmHs} \\
 & m_{hH}^2 \, = \,\frac{1}{8} \left(4 \cos 2 \beta  \left( \sqrt{2} v_s (a_\lambda + \lambda M) + 2 B_\mu \right) - 3 v^2 \sin 4 \beta 
   \left( \lambda^2 - 4 \tilde{g}^2 \right) + 4 \sin 2 \beta  (m_{H_d}^2 - m_{H_u}^2) \right)  \, .\label{MassMatrixElementsNMSSMdspmhH} 
\end{align}
\end{subequations}
For the scale-invariant NMSSM with superpotential \eqref{scaleinvNMSSM} (parameterised by the dimensionless couplings $\lambda, \kappa$), the matrix elements become 
\begin{subequations}\label{MassMatrixElementsNMSSMsi}
\begin{align}
& \begin{multlined}
\label{NMSSMPotentialParameters2}
m_h^2 \, = \, \frac{1}{8} \Bigl( 24 \tilde{g}^2 v^2 \cos^2 \hspace*{-0.035cm} 2\beta +6 \lambda^2 v^2 \sin^2 \hspace*{-0.035cm} 2\beta +8 m_{H_d}^2 \cos^2 \hspace*{-0.035cm} \beta  + 8 m_{H_u}^2 \sin^2\hspace*{-0.035cm} \beta-4 \sqrt{2} a_\lambda  v_s \sin 2 \beta  \\ 
- 4 \kappa  \lambda  v_s^2 \sin 2 \beta +4 \lambda^2 v_s^2 \Bigr) \end{multlined}\\ 
& \begin{multlined}
m_H^2 \, = \,  \frac{1}{8} \Bigl( 4 \sqrt{2} a_\lambda  v_s \sin 2 \beta - 3 v^2 \cos 4 \beta  \left(4 \tilde{g}^2 - \lambda^2 \right) + 4 \tilde{g}^2 v^2 + 8 m_{H_ d}^2 \sin^2 \hspace*{-0.035cm} \beta +8 m_{H_u}^2\cos^2  \hspace*{-0.035cm} \beta\\ 
   + \lambda^2 v^2 + 4 \kappa  \lambda  v_s^2 \sin 2 \beta +4 \lambda^2 v_s^2 \Bigr)
\end{multlined} \\
& m_s^2 \, = \, \sqrt{2} a_\kappa v_s + m_S^2 - \kappa  \lambda  v^2 \sin \beta  \,  \cos \beta + \frac{1}{2}\lambda^2 v^2 + 3 \kappa^2 v_s^2 \\ 
& m_{hH}^2 \, = \, \frac{1}{4} \left( \cos 2 \beta  \left(2 v_s \left(\sqrt{2} a_\lambda + \kappa \lambda  v_s \right) + 3 v^2 \sin 2 \beta  \left(4 \tilde{g}^2 - \lambda^2 \right)\right) + 2 \sin 2 \beta  (m_{H_d}^2 - m_{H_u}^2 )\right)\\
& m_{hs}^2 \, = \,  v \left( \lambda^2 v_s - \sin \beta  \,  \cos \beta  \left(\sqrt{2} a_\lambda + 2 \kappa  \lambda  v_s \right)\right) \\
& \label{NMSSMPotentialParameters9}
m_{Hs}^2 \, = \, \frac{v}{2}  \cos 2 \beta  \left(\sqrt{2} a_\lambda + 2 \kappa  \lambda  v_s \right) \, .
\end{align}
\end{subequations}

\section{Approximate mass and couplings of the Higgs}
\label{app:HiggsCouplings}

In this appendix, we will derive approximate expressions for the mass and couplings of the Higgs. We will present the derivation for the case of two Higgs doublets and one singlet as in the NMSSM. However, it applies to any type-II two-Higgs-doublet model and can be straightforwardly extended to an arbitrary number of additional Higgs singlets. The corresponding expressions for the MSSM can be obtained by decoupling the singlet. We rotate into the basis $(h,H,s)$ using eq.~\eqref{basischange1aNMSSM}. The couplings to SM fermions $f$ and SM vectors $V^\mu$ in this basis are given by
\begin{equation}\label{HiggsCouplingLagrangian}
\mathcal{L} \, \supset \, -\bar{f}f \,  (Y^h_f h +Y^H_f H)  -  g^h_{V} \, V_{\mu}V^{\mu} \, h \, . 
\end{equation} 
Note that the state $H$ does not couple to vectors (and the singlet $s$ of course couples to neither fermions nor vectors). Furthermore the state $h$ couples precisely like the SM Higgs. Accordingly $Y^h_f $ are just the SM Yukawa couplings and similarly $g^h_{V}$ are the SM Higgs couplings to vectors. For the Yukawa couplings $\smash{Y^H_{f_u}}$ and $\smash{Y^H_{f_d}}$ to respectively up-type and down-type fermions, on the other hand, we find (see e.g.~\cite{Gupta:2012fy})
\begin{equation}
Y^H_{f_u} \, = \, -\cot\beta  \, Y^h_{f_u}  \, ,\qquad  Y^H_{f_d} \, = \, \tan\beta \, Y^h_{f_d}   \, .
\end{equation}

We identify the Higgs (i.e.~the particle which was observed at the LHC) with the lightest $CP$-even state in the Higgs sector. In order to determine its mass and couplings, we need to diagonalize the mass matrix. 
In terms of the resulting mass eigenstates $h_{1,2,3}$, the (gauge eigenstates) $h,H,s$ are given by
\begin{equation}
h \, =\, \sum_{n=1}^{3} V_{hn} h_n\, , \qquad H \, = \, \sum_{n=1}^{3}V_{Hn} h_n\, , \qquad s \, = \, \sum_{n=1}^{3} V_{sn} h_n \, , \label{expansion}
\end{equation}
where $V$ is the unitary matrix such that $\smash{V^\dagger \mathcal{M}^2 V}$ is diagonal. Using these relations in eq.~\eqref{HiggsCouplingLagrangian}, the coupling ratios for the lightest $CP$-even state $h_1$ defined in eq.~\eqref{CouplingRatioDef} are given by
\begin{equation}\label{CouplingRatiosGen}
r_{f} \, =\, V_{h1} +\frac{Y^H_f}{Y^h_f} V_{H1} \, , \, \, \qquad r_V=V_{h1}  \, .
\end{equation}
Note that the Higgs coupling to vectors is always suppressed compared to the SM as $V_{h1} \leq 1$. This can be understood from the effect of the admixed states $H$ and $s$ on the wavefunction renormalization of the Higgs.

We are interested in the limit of weak mixing where the diagonal elements $m_H^2$ and $m_s^2$ of the mass matrix are much larger than all other matrix elements.  We can thus approximately diagonalize the mass matrix in the limit of large $m_H^2$ and $m_s^2$. In particular, this gives eq.~\eqref{HiggsMassNMSSM} for the mass of the lightest $CP$-even state $h_1$ (and eq.~\eqref{HiggsMassMSSM} after decoupling the singlet). For the mixing matrix elements $V_{i1}$ of the eigenstate $h_1$, we find
\begin{subequations}
\begin{align}
V_{h1}& \, \simeq \, 1-\frac{m_{hH}^4}{2m_H^4}-\frac{m_{hs}^4}{2m_s^4}\\
V_{H1} &\, \simeq \, -\frac{m_{hH}^2}{m_H^2}+\frac{m_{hs}^2 m_{Hs}^2}{m_H^2m_s^2}-\frac{m_{h}^2 m_{hH}^2}{m_H^4}\\
V_{s1}& \, \simeq \,  -\frac{m_{hs}^2}{m_s^2}+\frac{m_{hH}^2 m_{Hs}^2}{m_H^2m_s^2}-\frac{m_{h}^2 m_{hs}^2}{m_s^4} \, .
\end{align}
\end{subequations}
Plugging these relations into eq.~\eqref{CouplingRatiosGen}, we obtain eq.~\eqref{rurdrvNMSSM} (and eq.~\eqref{rurdrvMSSM} after decoupling the singlet).

We can express $\tan\beta$ in terms of the Higgs couplings. Combining eqs.~\eqref{CouplingRatiosGen}, we find
\begin{equation}
r_u - r_V \,  \simeq \, -\cot \beta \, V_{H1}\;, \qquad r_d - r_V \, \simeq \, \tan\beta \, V_{H1}\;.
\end{equation}
If the Higgs has a non-vanishing admixture of $H$ ($V_{H1}\neq0$), we then obtain the simple relation
\begin{equation}
\tan^2 \hspace*{-0.035cm}  \beta \,  \simeq \, \frac{r_d-r_V}{r_V-r_u}\;.
\end{equation}
This result is general for any type-II two-Higgs-doublet model including an arbitrary number of singlets. However, it is difficult to use this relation to measure $\tan\beta$ because of the large experimental uncertainty in $r_u$.

\section{Expressions used in the evaluation of the fine-tuning measure}
\label{Finetuningexpressions}
In order to evaluate the fine-tuning measure \eqref{finetuningdef}, we need to calculate the logarithmic derivatives of the Higgs vev $v$ with respect to the input parameters $\xi$. For completeness, we present the expressions used in this calculation for both the MSSM and NMSSM.
\subsection{MSSM}
 In a general basis $(h_1,h_2)$ for the $CP$-even Higgs fields, the minimisation conditions for the potential read
\begin{equation}\label{minconditions}
\ell_i \, \equiv \, \left.\frac{\partial V}{\partial h_i}\right|_{\mathrm{min}} \, = \, 0 \, .
\end{equation}
In particular for the basis $(h,H)$ and using the fact that the minimum in this basis is by construction at $\langle h \rangle  = v$ and  $\langle H \rangle = 0$, we find
\begin{subequations}\label{minconditionsMSSM}
\begin{align}
& \ell_h \,  = \, v \, \bigl( m_{H_d}^2 \cos^2 \hspace*{-0.035cm}  \beta+ m_{H_u}^2 \sin^2\hspace*{-0.035cm} \beta + \tilde{g}^2 v^2 \cos^2 \hspace*{-0.035cm} 2\beta + \mu^2 - B_\mu \sin 2 \beta \bigr) \, = \, 0 \label{ellmathcalh}\\
 & \ell_H  \,  = \, \frac{v}{2} \, \bigl((m_{H_d}^2 - m_{H_u}^2 + 2 \tilde{g}^2 v^2 \cos 2 \beta) \sin 2 \beta +2 B_\mu \cos 2 \beta\bigr) \, = \, 0 \, .\label{ellmathcalH}
\end{align}
\end{subequations}
These equations can be used to solve for $v$ and $\tan \beta$ for given parameters $\{m_{H_u}^2, m_{H_d}^2, \mu, B_\mu, \tilde{g} \}$ or, alternatively, to fix the soft masses $m_{H_u}^2$ and $m_{H_d}^2$ for given $\{v, \tan \beta, \mu, B_\mu, \tilde{g}\}$.
Taking the total derivative of the minimisation conditions with respect to the input parameters $\xi$, we find
\begin{equation}\label{eq:derFT}
0 \, = \, \frac{d \ell_i}{d\xi} \, = \, \frac{\partial \ell_i}{\partial \langle h_j \rangle} \frac{d \langle h_j \rangle}{d \xi}+\frac{\partial \ell_i}{\partial \xi} \, = \, \left.\frac{\partial^2 V }{\partial h_i \partial h_j }\right|_{\mathrm{min}} \hspace*{-0.1cm} \frac{d \langle h_j \rangle}{d \xi}+\frac{\partial \ell_i}{\partial \xi} \, = \, \left(\mathcal{M}^2\right)_{ij}  \frac{d \langle h_j \rangle }{d \xi}+\frac{\partial \ell_i}{\partial \xi} \, .
\end{equation}
This can be solved for the $d \langle h_j \rangle / d \xi$. In the field basis $(h,H)$, this gives an expression for $d v / d \xi$ which in combination with the definition \eqref{finetuningdef} of the fine-tuning measure yields \eqref{finetuning2}.

In the basis $ \{ v, \tan \beta, \mu, m_h, m_H \}$ of input parameters, the resulting expression for the fine-tuning measure reads
\begin{equation}
\begin{split}
\Sigma & =\frac{1}{4} \left( \left(  \cos^4 \hspace*{-0.035cm} \beta \sec^6 \hspace*{-0.035cm} 2 \beta \,  \left( m_h^2 - \cos 2 \beta  \left(m_h^2+m_H^2 \right)\right)^2 \left(4 \mu^2 + m_h^2
 \cos 6 \beta -2 \cos 4 \beta  \left( -2 \mu^2 + m_h^2 + m_H^2 \right) 
 \right.  \right.  \right.\\ &  \left.  \left.  \left.  
 + \,  3 \cos 2 \beta  \left(m_h^2 + m_H^2 \right) + 2
   m_h^2+ m_H^2 \cos 6 \beta - 2 m_H^2 \right)^2 + \sin^4 \hspace*{-0.035cm} \beta \,  \sec^6 \hspace*{-0.035cm} 2 \beta  \left(\cos 2 \beta 
   \left(m_h^2 + m_H^2 \right) + m_h^2 \right)^2
  \right. \right.  \\ & \left.  \left.  
    \left( -4 \mu^2 + m_h^2 \cos 6 \beta + 2 \cos 4 \beta  \left( -2 \mu^2 + m_h^2+ m_H^2 \right) + 3 \cos 2 \beta  \left(m_h^2 + m_H^2 \right) - 2 m_h^2 + m_H^2 \cos 6 \beta 
     \right. \right. \right. \\ & \left. \left. \left.
     +\, 2 m_H^2 \right)^2  + 4 \tan^4 \hspace*{-0.035cm} 2 \beta  \left(m_h^2 + m_H^2 \right)^2 \left(\cos 4 \beta 
   \left(m_h^2 + m_H^2 \right) - m_h^2 + m_H^2 \right)^2 + 16 m_h^4 \sec^4 \hspace*{-0.035cm} 2 \beta  
   \right. \right. \\ & \left. \left.    
   \left(\cos 4 \beta \left(m_h^2+mH^2 \right) 
   - m_h^2 + m_H^2 \right)^2 + 256 \mu^4 m_H^4 \right)  /   \left(m_h^2 m_H^2 - m_h^4 \tan^2 \hspace*{-0.035cm} 2 \beta \right)^2 \right)^{1/2}
\label{SigmaMSSM}
\end{split}
\end{equation}
In appropriate limits, this simplifies to the expressions in eqs.~\eqref{fine-tuning-limit} and \eqref{fine-tuning-large-tanbeta}.

\subsection{NMSSM}
We first extend eq.~\eqref{eq:derFT} to the case of three fields. Solving for the derivatives $d v / d \xi$, we then find
\be
\frac{d v }{d \xi} \, = \, \frac{ \ell_h' m_{Hs}^4 - \ell_h' m_H^2 m_s^2 + \ell_H' m_{hH}^2 m_s^2- \ell_H' m_{hs}^2
   m_{ Hs}^2 + \ell_s' m_H^2 m_{hs}^2 - \ell_s' m_{hH}^2 m_{Hs}^2}{\det \mathcal{M}^2} \, , \label{dvdxiNMSSM}
\ee
where $\ell_s' \equiv d \ell_s / d \xi$ etc. The $\ell_{h,H,s}$ are defined as in eq.~\eqref{minconditions}. For the NMSSM with mass terms in the superpotential (given in \eqref{spmt}), this gives
\begin{subequations}  \label{minconditionsNMSSMdsp}
\begin{align}
& \begin{multlined}
\ell_h \, = \,  \frac{v}{8}  \left(  v^2 \cos 4 \beta  \left(4 \tilde{g}^2 - \lambda^2 \right) -4 \sin 2 \beta  (\sqrt{2} v_s (a_\lambda  + \lambda M)+ 2 B_\mu ) + 4 \tilde{g}^2 v^2   + 4 \cos 2 \beta  ( m_{H_d}^2 - m_{H_u}^2 ) \right. \\ \left.  + 4 m_{H_d}^2 + 4 m_{H_u}^2+ \lambda^2 v^2 + 4 v_s^2 \lambda^2 + 8 \mu^2 + 8 \sqrt{2} v_s \lambda \mu   \right) 
\end{multlined} \\
& \ell_H \, = \,\frac{v}{8} \left(  4 \cos 2 \beta \,  (2 B_\mu + \sqrt{2} v_s (a_\lambda + \lambda M) ) - v^2 \sin 4 \beta  \left( \lambda^2 - 4
   \tilde{g}^2 \right) + 4 \sin 2 \beta  (m_{H_d}^2 - m_{H_u}^2 ) \right) \\
& \ell_s \, = \, \frac{1}{2} \left(  2 M^2 v_s + 2 m_S^2 v_s + \lambda  v^2 ( \sqrt{2} \mu + \lambda v_s)- \sqrt{2} v^2 \sin \beta  \cos \beta  (a_\lambda + \lambda  M) \right) \, .
\end{align}
\end{subequations}
For the NMSSM with a scale-invariant superpotential (given in eq.~\eqref{scaleinvNMSSM}), on the other hand, we find
\begin{subequations} \label{minconditionsNMSSMsi}
\begin{align}
&\begin{multlined}[t][10.5cm]
\label{NMSSMPotentialParameters1}
\ell_h \, = \, \frac{v}{8}  \Bigl(v^2 \cos 4 \beta  \left(4 \tilde{g}^2 - \lambda^2 \right) - 4 \sqrt{2} a_\lambda  v_s \sin 2 \beta  + 4
   \tilde{g}^2 v^2 + 4 \cos 2 \beta  (m_{H_d}^2 - m_{H_u}^2 ) + 4 m_{H_u}^2  \\   +  4 m_{H_d }^2 + \lambda^2 v^2 - 4 \kappa  \lambda  v_s^2 \sin 2 \beta + 4 \lambda^2 v_s^2 \Bigr)   
   \end{multlined} \\
& \ell_H \, = \,\frac{v}{8} \left(\cos 2 \beta  \left(4 v_s \left(\sqrt{2} a_\lambda + \kappa  \lambda  v_s \right) + 2 v^2 \sin 2 \beta \left(4 \tilde{g}^2 - \lambda^2 \right)\right) + 4 \sin 2 \beta  (m_{H_d}^2  -m_{H_u}^2 )\right)\\
& \ell_s \, = \,   \frac{v_s}{2} \left(\sqrt{2} a_\kappa  v_s + 2 m_S^2 + \lambda^2 v^2 + 2 \kappa^2 v_s^2 \right) - \frac{1}{4} v^2
   \sin 2 \beta  \left(\sqrt{2} a_\lambda + 2 \kappa  \lambda  v_s \right) \, .
\end{align}
\end{subequations}

\bibliography{FineTunedHiggs}

\end{document}